\documentclass[twocolumn,showpacs,preprintnumbers,superscriptaddress,amsmath,amssymb,aps,pre]{revtex4-1}

\usepackage{graphicx}
\usepackage{dcolumn}
\usepackage{bm}
\usepackage{color}
\usepackage{longtable}

\begin{document}


\title{Random matrix approach to the dynamics of stock inventory variations}

\author{Wei-Xing Zhou}
 \email{wxzhou@ecust.edu.cn}
 \affiliation{School of Business, East China University of Science and Technology, Shanghai 200237, China} %
 \affiliation{School of Science, East China University of Science and Technology, Shanghai 200237, China} %
 \affiliation{Research Center for Econophysics, East China University of Science and Technology, Shanghai 200237, China} %
\author{Guo-Hua Mu}
 \affiliation{School of Business, East China University of Science and Technology, Shanghai 200237, China} %
 \affiliation{School of Science, East China University of Science and Technology, Shanghai 200237, China} %
 \affiliation{Research Center for Econophysics, East China University of Science and Technology, Shanghai 200237, China} %
\author{J{\'a}nos Kert{\'e}sz}
 \email{kertesz@phy.bme.hu}
 \affiliation{Department of Theoretical Physics, Budapest University of Technology and Economics, Budapest, Hungary} %
 \affiliation{Laboratory of Computational Engineering, Helsinki University of Technology, Espoo, Finland} %

\date{\today}

\begin{abstract}
Investors trade stocks based on diverse strategies trying to beat the market and gain access returns, whose stock inventories change accordingly. The dynamics of inventory variations thus contain rich information about the trading behaviors of investors and have crucial influence on price fluctuations. We study the cross-correlation matrix $C_{ij}$ of inventory variations of the most active individual and institutional investors in an emerging market to understand the dynamics of inventory variations. We find that the distribution of cross-correlation coefficient $C_{ij}$ has a power-law form in the bulk followed by exponential tails and there are more positive coefficients than negative ones. In addition, it is more possible that two individuals or two institutions have stronger inventory variation correlation than one individual and one institution. We find that the largest and the second largest eigenvalues ($\lambda_1$ and $\lambda_2$) of the correlation matrix cannot be explained by the random matrix theory and the projection of inventory variations on the first eigenvector $u(\lambda_1)$ are linearly correlated with stock returns, where individual investors play a dominating role. The investors are classified into three categories based on the cross-correlation coefficients $C_{VR}$ between inventory variations and stock returns. Half individuals are reversing investors who exhibit evident buy and sell herding behaviors, while 6\% individuals are trending investors. For institutions, only 10\% and 8\% investors are trending and reversing investors. A strong Granger causality is unveiled from stock returns to inventory variations, which means that a large proportion of individuals hold the reversing trading strategy and a small part of individuals hold the trending strategy. Comparing with the case of Spanish market, Chinese investors exhibit common and market-specific behaviors. Our empirical findings have scientific significance in the understanding of investors' trading behaviors and in the construction of agent-based models for stock markets.
\end{abstract}

\pacs{89.65.Gh, 89.75.Da, 02.10.Yn, 05.45.Tp}

\maketitle


\section{Introduction}
\label{intro}

Stock markets are complex systems, whose elements are heterogenous individual and institutional investors interacting with each other by stock exchanges \cite{Bouchaud-2008-Nature,Lux-Westerhoff-2009-NPhys,Farmer-Foley-2009-Nature,Schweitzer-Fagiolo-Sornette-VegaRedondo-Vespignani-White-2009-Science}. Stock price fluctuates due to investors' trading activities and the cross-sectional relation between investors' stock inventory variations and stock returns have attracted much attention \cite{Lillo-Moro-Vaglica-Mantegna-2008-NJP}. The huge literature falls into three groups to study the relation between past returns and inventory variations, to investigate the contemporaneous relation between inventory variations and stock returns, and to analysis return predictability of inventory variations \cite{Griffin-Harris-Topaloglu-2003-JF}. The main findings are that institutions are trending investors adopting the momentum trading strategy \cite{Griffin-Harris-Topaloglu-2003-JF,Choe-Kho-Stulz-1999-JFE,Grinblatt-Keloharju-2000-JFE}, while individuals are reversing investors who buy previous losers and sell previous winners \cite{Griffin-Harris-Topaloglu-2003-JF,Odean-1998a-JF,Barber-Odean-2000-JF,Grinblatt-Keloharju-2000-JFE,Barber-Lee-Liu-Odean-2009-RFS}, and stock returns lead inventory variations but not {\em{vice versa}} \cite{Choe-Kho-Stulz-1999-JFE,Griffin-Harris-Topaloglu-2003-JF,Grinblatt-Keloharju-2000-JFE,Lillo-Moro-Vaglica-Mantegna-2008-NJP}.

However, there is evidence showing different trading patterns. Lillo et al investigated the trading behaviors of about 80 firms that were members of the Spanish Stock Exchange and found that there were more reversing firms than trending firms \cite{Lillo-Moro-Vaglica-Mantegna-2008-NJP}. They also found that the largest eigenvalue of the correlation matrix of inventory variations cannot be explained by the random matrix theory and its eigenvector contains information of stock price fluctuations. Both buying and selling herding behaviors have been observed for trending and reversing firms.

In this work, we perform a similar analysis as in Ref.~\cite{Lillo-Moro-Vaglica-Mantegna-2008-NJP} based on the trading records of Chinese investors in the Shenzhen Stock Exchange. Different from the Spanish case, our data set contains both individual and institutional investors, which allows us to observe interesting investor behaviors. Our analysis starts from the perspective of random matrix theory, which has been extensively used to investigate the cross-correlations of financial returns in different stock markets \cite{Laloux-Cizean-Bouchaud-Potters-1999-PRL,Plerou-Gopikrishnan-Rosenow-Amaral-Stanley-1999-PRL,Plerou-Gopikrishnan-Rosenow-Amaral-Guhr-Stanley-2002-PRE}. However, very few studies have been conducted on the Chinese stocks \cite{Shen-Zheng-2009a-EPL} and, to our knowledge, there is no research reported on the dynamics of inventory variations of Chinese investors. Alternatively, there are studies on Chinese equities at the transaction and trader level from the complex network perspective \cite{Jiang-Zhou-2010-PA,Wang-Zhou-2011-PA,Sun-Cheng-Shen-Wang-2011-PA,Wang-Zhou-Guan-2011-XXX}.

This paper is organized as follows. Section \ref{S1:Data} describes the data and the method to construct the time series of investors' inventory variations. Section \ref{S1:Cij} studies the statistical properties of the elements, eigenvalues and eigenvectors of the correlation matrix of inventory variations. Section \ref{S1:CVR} investigates the contemporaneous and lagged cross-correlation between inventors' inventory variations and stock returns to divide investors into three categories and their herding behaviors. Section \ref{S1:Summary} summarizes our findings.

\section{Data}
\label{S1:Data}

We analyze 39 stocks actively traded on the Shenzhen Stock Exchange in 2003. The data base contains all the information needed for the analysis in this work. For each transaction $i$ of a given stock, the data record the identities of the buyer and seller, the types (individual or institution) of the two traders, the price $p_i$ and the size $q_i$ of the trade, and the time stamp. Therefore, the trading history of each investor is known. For each stock, we identify active traders who had more than 150 transactions, amount to about three transactions per week. If the number of active traders of a stock is less than 120, we exclude it from analysis. In this way, we have 15 stocks for analysis.

\begin{table*}[htb]
 \centering
 \caption{\label{TB:BasicStat} Basic statistics of the investigated stocks. The first column is the stock code, which is the unique identity of each stock. The second and third columns presents investor-averaged total inventory variation $\langle{\sum_t{v}}\rangle_i$ and average absolute variation $\langle{\langle{|v|}\rangle}_t\rangle_i$. The fourth to eighth columns gives the number of investors $N$, the number of trending investors $N^{\rm{tr}}$, the number of reversing investors $N^{\rm{re}}$, the number of uncategorized investors $N^{\rm{un}}$, and the slope of the factor versus stock return $k$. The variables in the ninth to thirteenth columns are the same as in the five ``all investors'' columns but for individual investors and the last five columns are for institutional investors. Each value in the last row gives the sum of the numbers in the same column.}
 \medskip
 \begin{tabular}{cccccccccccccccccccccccccccccccccccccccccccccccccc}
  \hline  \hline
  &&&&& \multicolumn{5}{c}{All investors}&&& \multicolumn{5}{c}{Individuals}&&& \multicolumn{5}{c}{Institutions}       \\
  \cline{6-10}\cline{13-17}\cline{20-24}
  Code & $\langle{\sum_t{v}}\rangle_i$ & $\langle{\langle{|v|}\rangle}_t\rangle_i$ &&&
  $N$ & $N_{\rm{tr}}$ & $N_{\rm{re}}$ & $N_{\rm{un}}$ & $k$ &&&
  $N_{\rm{ind}}$ & $N_{\rm{ind}}^{\rm{tr}}$ & $N_{\rm{ind}}^{\rm{re}}$ & $N_{\rm{ind}}^{\rm{un}}$ & $k_{\rm{ind}}$ &&&
  $N_{\rm{ins}}$ & $N_{\rm{ins}}^{\rm{tr}}$ & $N_{\rm{ins}}^{\rm{re}}$ & $N_{\rm{ins}}^{\rm{un}}$ & $k_{\rm{ins}}$
  \\\hline
     000001 & $-2.99\times10^6$ & $1.12\times10^5$ &&& 80 & 7 & 41 & 32 & 0.83 &&& 61 & 4 & 39 & 18 & 0.83 &&& 19 & 3 & 2 & 14 & 0.04 \\
     000002 & $1.01\times10^6$ & $1.46\times10^5$ &&& 80 & 6 & 29 & 45 & 0.49 &&& 42 & 2 & 26 & 14 & 0.52 &&& 38 & 4 & 3 & 31 & 0.06 \\
     000012 & $-1.64\times10^6$ & $7.88\times10^4$ &&& 81 & 5 & 20 & 56 & 0.19 &&& 78 & 5 & 20 & 53 & 0.18 &&& 3 & 0 & 0 & 3 & 0.06 \\
     000021 & $5.09\times10^5$ & $4.35\times10^4$ &&& 81 & 2 & 43 & 36 & 0.78 &&& 64 & 2 & 43 & 20 & 0.78 &&& 17 & 2 & 0 & 16 & 0.04 \\
     000063 & $1.46\times10^7$ & $2.46\times10^5$ &&& 80 & 9 & 20 & 51 & 0.13 &&& 20 & 2 & 13 & 5 & 0.71 &&& 60 & 7 & 7 & 46 & 0.05 \\
     000488 & $1.35\times10^6$ & $8.92\times10^4$ &&& 80 & 5 & 5 & 70 & 0.08 &&& 69 & 2 & 5 & 63 & 0.06 &&& 11 & 4 & 0 & 7 & 0.06 \\
     000550 & $4.20\times10^6$ & $7.53\times10^4$ &&& 81 & 2 & 37 & 42 & 0.18 &&& 45 & 0 & 35 & 10 & 0.18 &&& 36 & 2 & 2 & 32 & 0.06 \\
     000625 & $1.87\times10^6$ & $1.22\times10^5$ &&& 80 & 10 & 26 & 44 & 0.71 &&& 62 & 8 & 24 & 30 & 0.69 &&& 18 & 2 & 2 & 14 & 0.05 \\
     000800 & $2.53\times10^6$ & $2.56\times10^5$ &&& 80 & 6 & 19 & 55 & 0.30 &&& 31 & 2 & 17 & 13 & 0.31 &&& 49 & 5 & 2 & 42 & 0.06 \\
     000825 & $6.33\times10^6$ & $9.38\times10^4$ &&& 80 & 5 & 38 & 37 & 0.69 &&& 50 & 3 & 37 & 10 & 0.70 &&& 30 & 2 & 2 & 27 & 0.05 \\
     000839 & $8.13\times10^5$ & $6.67\times10^4$ &&& 80 & 2 & 40 & 38 & 0.84 &&& 60 & 2 & 38 & 20 & 0.84 &&& 20 & 0 & 2 & 18 & 0.04 \\
     000858 & $1.75\times10^6$ & $1.66\times10^5$ &&& 82 & 4 & 21 & 57 & 0.33 &&& 31 & 0 & 19 & 12 & 0.67 &&& 51 & 4 & 2 & 45 & 0.05 \\
     000898 & $6.26\times10^6$ & $1.20\times10^5$ &&& 83 & 6 & 32 & 45 & 0.79 &&& 46 & 0 & 31 & 15 & 0.80 &&& 37 & 6 & 2 & 30 & 0.04 \\
     200488 & $3.59\times10^6$ & $5.05\times10^4$ &&& 80 & 9 & 30 & 41 & 0.65 &&& 53 & 7 & 25 & 21 & 0.66 &&& 27 & 2 & 5 & 20 & 0.05 \\
     200625 & $6.27\times10^6$ & $1.26\times10^5$ &&& 83 & 9 & 14 & 60 & 0.50 &&& 46 & 5 & 9 & 32 & 0.58 &&& 37 & 4 & 5 & 28 & 0.05 \\
     $\sum$ & - & - &&& 1211 & 87 & 415 & 709 &- &&& 758 & 44 & 381 & 336 & - &&& 453 & 47 & 36 & 373 & - \\
  \hline  \hline
\end{tabular}
\end{table*}

Following Ref.~\cite{Lillo-Moro-Vaglica-Mantegna-2008-NJP}, we investigate the dynamics of the inventory variation of the most active investors who executed more than 120 transactions for each stock. Although the trading period of each day consists of call auction and continuous auction, their behaviors are different in many aspects and are usually studied separately \cite{Gu-Chen-Zhou-2008b-PA,Gu-Ren-Ni-Chen-Zhou-2010-PA}. We stress that all the transactions in both call auction and continuous auction are included in our investigation. The daily inventory variation of an investor $i$ trading a given stock on day $t$ is defined as follows
\begin{equation}
 v_{i}(t) = \sum^{+}p_{i}(t)q_{i}(t)-\sum^{-}p_{i}(t)q_{i}(t),
 \label{Eq:v}
\end{equation}
where $\sum^{+}p_{i}(t)q_{i}(t)$ is the total buy quantity on trading day $t$ and $\sum^{-}p_{i}(t)q_{i}(t)$ is the total sell quantity in the same day. The basic statistics of the 80 most active traders and the resultant inventory variations are given in Table \ref{TB:BasicStat}.

\section{Statistics of correlation matrix between two time series of inventory variations}
\label{S1:Cij}

\subsection{Distributions of cross-correlation coefficients}
\label{S2:PDF:Cij}

The empirical correlation matrix $C$ is constructed from the time series of inventory variation $v_{i}(t)$ of the investigated stock, defined as
\begin{equation}
 C_{ij}=\frac{\langle(v_{i}-\langle v_{i}\rangle)(v_{j}-\langle v_{j}\rangle)\rangle}{\sigma_{i}\sigma_{j}}.
 \label{Eq:Cor1}
\end{equation}
Since the results for individual stocks are quantitatively similar, we put the cross-correlation coefficients of the 15 stocks into one sample. We find that the mean value is $\langle{C_{ij}}\rangle=0.02$ for the real data and $0$ for the shuffled data. When the types of investors are taken into account, the mean value of the cross-correlation coefficients is $\langle{C_{ij}}\rangle=0.048$ (shuffled: $-0.001$) for both $i$ and $j$ being individuals, $\langle{C_{ij}}\rangle=0.014$ (shuffled: 0) for both $i$ and $j$ being institutions, and $\langle{C_{ij}}\rangle=-0.008$ (shuffled: $-0.001$) for $i$ being individual and $j$ being institution.

\begin{figure}[htb]
  \centering
  \includegraphics[width=7cm]{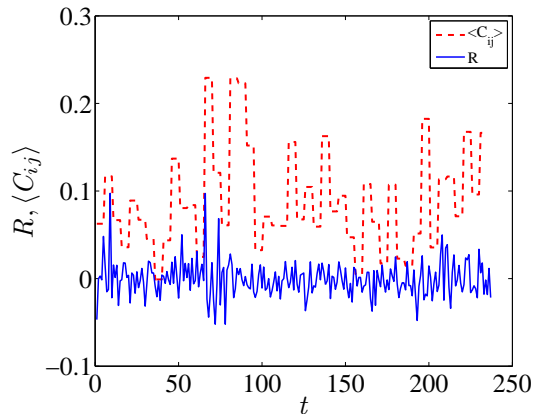}
  \caption{\label{Fig:Cij:R:000001} (Color online) Evolution of the 5-day average cross-correlation coefficient $\langle C_{ij}\rangle$ and the daily return $R$.}
\end{figure}

Figure \ref{Fig:Cij:R:000001} plots the daily returns of stock 000001 and the sliding average values of the correlation coefficients $\langle C_{ij}\rangle$ for comparison. We observe that large values $\langle C_{ij}\rangle$ appear during periods of large price fluctuations by and large, which is reminiscent of the similar result for cross-correlations of financial returns \cite{Plerou-Gopikrishnan-Rosenow-Amaral-Guhr-Stanley-2002-PRE}. However, the short time period of our data sample does not allow us to reach a decisive conclusion. There are also less volatile time periods with large $\langle C_{ij}\rangle$. The situation is quite similar for other stocks.

\begin{figure}[htb]
\centering
\includegraphics[width=6cm]{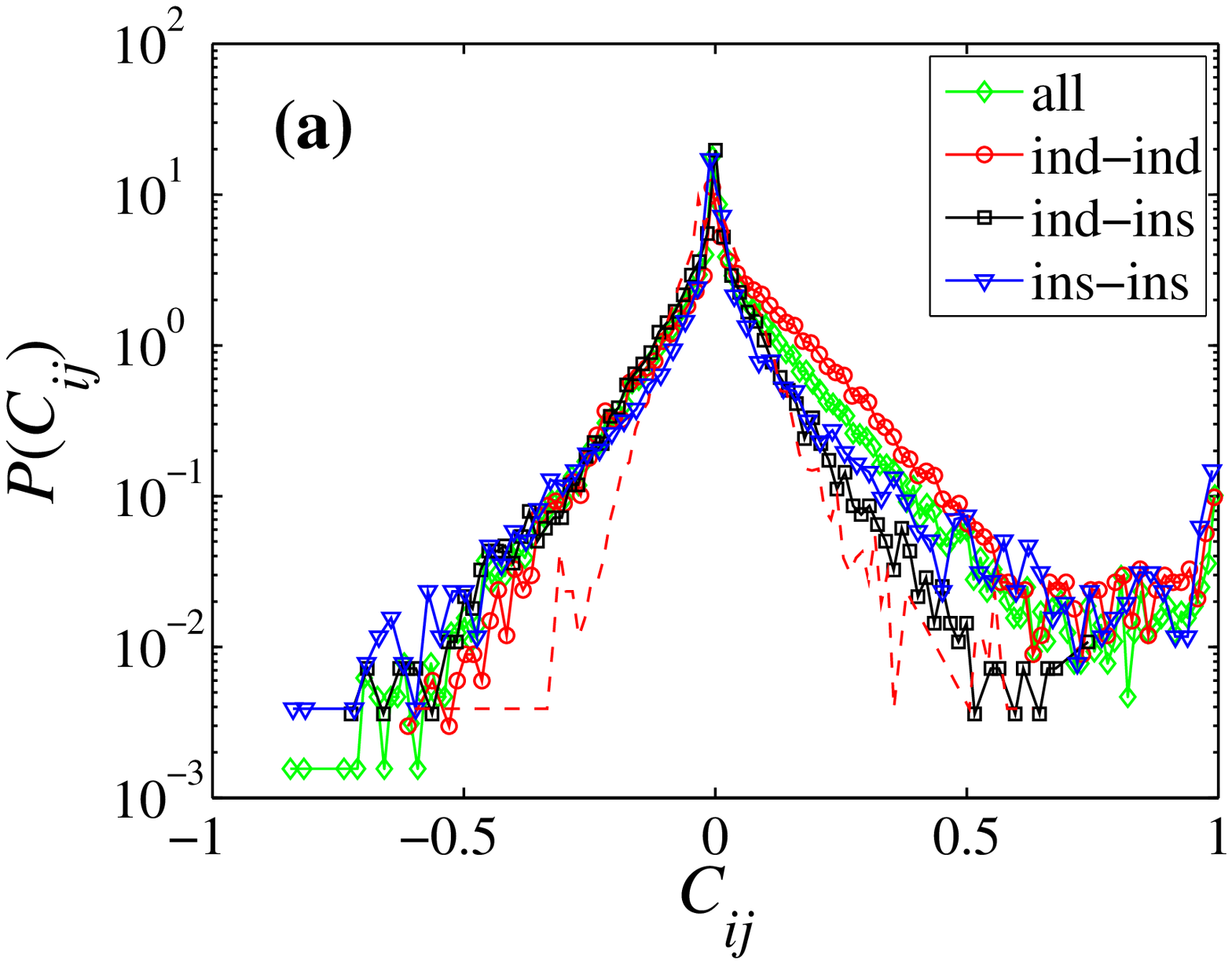}\hspace{3mm}
\includegraphics[width=6cm]{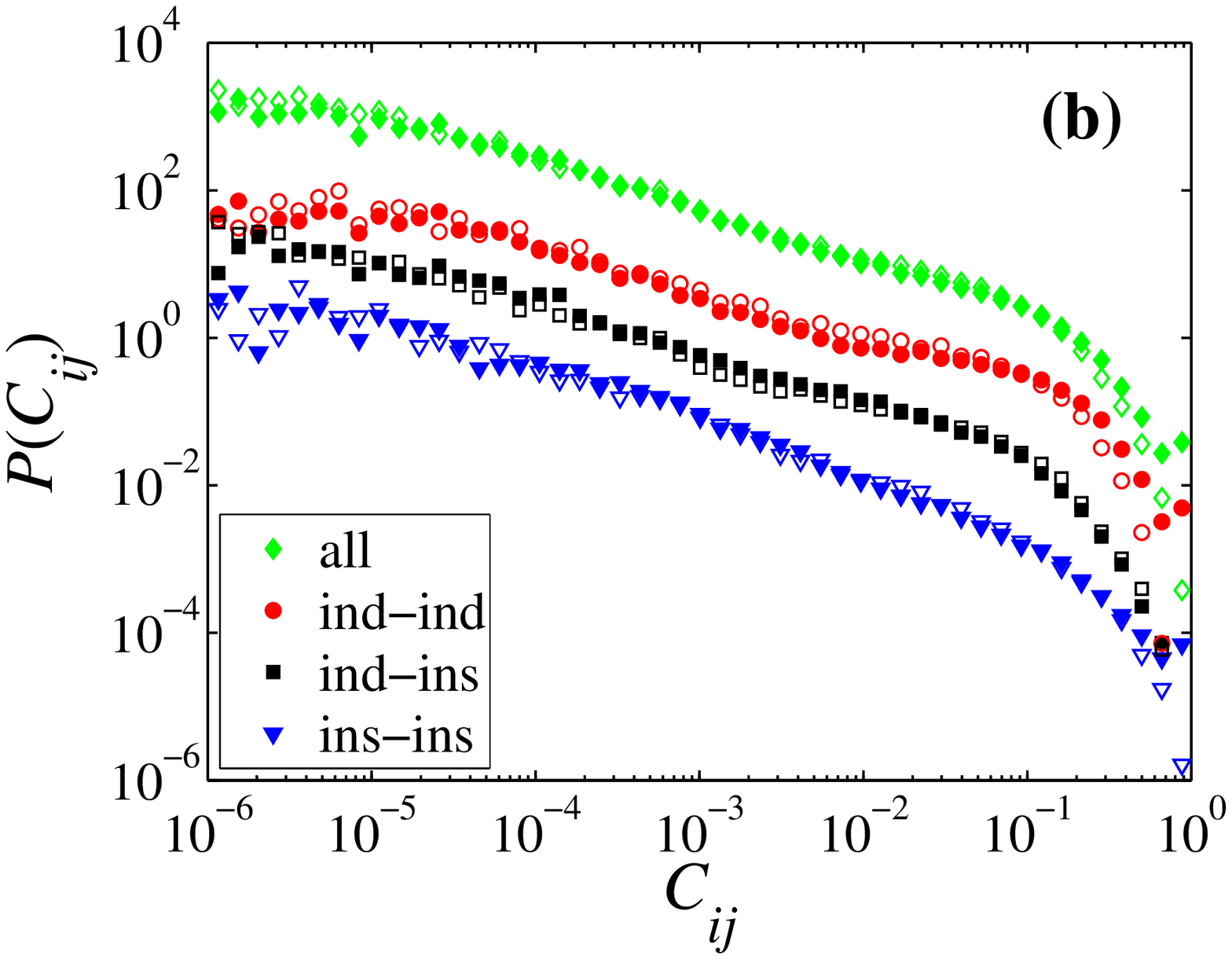}
\includegraphics[width=6cm]{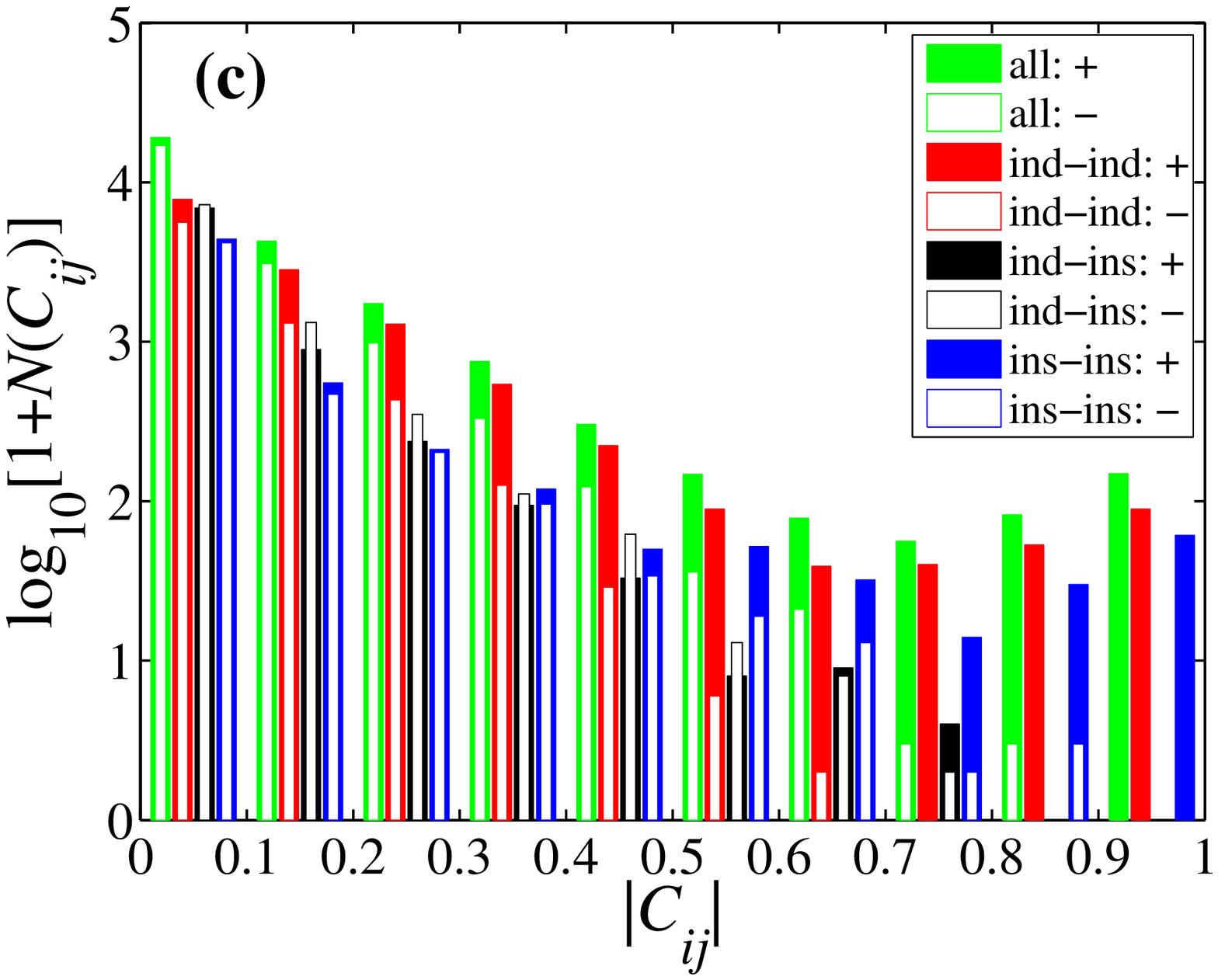}
\caption{\label{Fig:PDF:Cij} (Color online) Empirical distributions of the cross-correlation coefficients for all the 15 stocks. (a) Log-linear plot of $P(C_{ij})$ for cross-correlations between any two investors, any two individuals, any one individual and one institution, and any two institutions, which shows exponential forms when $0.1<|C|\leq0.6$. The dashed line corresponds to the result of shuffled data. (b) Log-log plot of $P(C_{ij})$, which shows power-law forms when $10^{-5}<|C|\leq0.01$. (c) Comparison of occurrence numbers of positive and negative cross-correlations. The ordinate gives $\log_{10}[1+N(C_{ij})]$ rather than $\log_{10}[N(C_{ij})]$ for better presentation.}
\end{figure}

Figure \ref{Fig:PDF:Cij}(a) shows the empirical probability distributions of $C_{ij}$ which is calculated using daily inventory variation. The four curves with different markers correspond to $C_{ij}$, $C_{\rm{ind,ind}}$, $C_{\rm{ind,ins}}$, and $C_{\rm{ins,ins}}$, respectively. It is found that most coefficients are small and the tails are exponentials:
\begin{equation}
 P(C) \propto \left\{
 \begin{array}{cccc}
  e^{-\lambda_-C}, & -0.6<C\leq-0.1\\
  e^{-\lambda_+C}, & 0.1<C\leq0.6
 \end{array}
 \right.
 \label{Eq:PDF:Cij:Exp}
\end{equation}
where $\lambda_+=8.8\pm0.2$ and $\lambda_-=11.1\pm0.3$ for $C_{ij}$, $\lambda_+=8.9\pm0.2$ and $\lambda_-=12.5\pm0.4$ for $C_{\rm{ind,ind}}$, $\lambda_+=10.8\pm0.4$ and $\lambda_-=10.5\pm0.4$ for $C_{\rm{ind,ins}}$, and $\lambda_+=6.6\pm0.5$ and $\lambda_-=8.7\pm0.5$ for $C_{\rm{ins,ins}}$, respectively. We find that there are more positive cross-correlation coefficients ($\lambda_+<\lambda_-$) when both investors are individuals or institutions. In contrast, the distribution is symmetric ($\lambda_+\approx\lambda_-$) when one investor is an individual while the other is an institution. This finding implies that herding behaviors are more like to occur among the same type of investors and individuals have larger probability to herd than institutions. We shuffle the original time series and perform the same analysis. The resulting distributions collapse onto a single curve, which has an exponential form
\begin{equation}
 P(C) = \lambda_{\rm{shuf}} e^{-\lambda_{\rm{shuf}}C}
 \label{Eq:PDF:Cij:Shuffle}
\end{equation}
where the parameter $\lambda_{\rm{shuf}}=23.3$ is determined using robust regression \cite{Holland-Welsch-1977-CStm,Street-Carroll-Ruppert-1988-AS}. It is not surprising that real data have higher cross-correlations than the shuffled data, which is confirmed by $\lambda_{\pm}<\lambda_{\rm{shuf}}$. This exponential distribution is different from the Gaussian distribution for the shuffled data of financial returns \cite{Plerou-Gopikrishnan-Rosenow-Amaral-Guhr-Stanley-2002-PRE}.

Figure \ref{Fig:PDF:Cij}(b) plots the distributions in double logarithmic coordinates where the negative parts are reflected to the right with respect to $C_{ij}=0$. Nice power laws are observed spanning over three orders of magnitude:
\begin{equation}
 P(C) \propto \left\{
 \begin{array}{cccc}
  (-C)^{-\gamma_-}, & -0.01<C\leq-10^{-5}\\
     C^{-\gamma_+}, & 10^{-5}<C\leq0.01
 \end{array}
 \right.
 \label{Eq:PDF:Cij:PL}
\end{equation}
where $\gamma_+=0.69\pm0.01$ and $\gamma_-=0.69\pm0.01$ for $C_{ij}$, $\gamma_+=0.67\pm0.02$ and $\gamma_-=0.62\pm0.02$ for $C_{\rm{ind,ind}}$, $\gamma_+=0.67\pm0.02$ and $\gamma_-=0.70\pm0.02$ for $C_{\rm{ind,ins}}$, and $\gamma_+=0.72\pm0.02$ and $\gamma_-=0.73\pm0.02$ for $C_{\rm{ins,ins}}$, respectively. It is found that $\gamma_-\approx\gamma_+$ and all the power-law exponents are close to each other. An intriguing feature is that the distributions of $C_{ij}$, $C_{\rm{ind,ind}}$ and $C_{\rm{ind,ins}}$ exhibit an evident bimodal behavior, which is reminiscent of the distributions of waiting times and interevent times of human short message communication \cite{Wu-Zhou-Xiao-Kurths-Schellnhuber-2010-PNAS}. Certainly, the underlying mechanisms are different and the factors causing the bimodal distribution of the cross-correlations are unclear.

It is natural that we are more interested in large cross-correlations. The preceding discussions focus on the cross-correlations not larger than 0.6. As shown in Fig.~\ref{Fig:PDF:Cij}(a), there are pairs of inventory variation time series that have very large cross-correlations that look like outliers. To have a better visibility, we plot in Fig.~\ref{Fig:PDF:Cij}(c) the numbers of occurrences of positive and negative cross-correlations in 10 nonoverlapping intervals for the four types of pairs. It is shown that $N(C>0)>N(C<0)$ in all intervals for $C=C_{ij}$, $C_{\rm{ind,ind}}$ and $C_{\rm{ind,ins}}$. In contrast, $N(C{\rm{ind,ins}}>0)<N(C{\rm{ind,ins}}<0)$ when $C{\rm{ind,ins}}<0.6$ and $N(C{\rm{ind,ins}}>0)>N(C{\rm{ind,ins}}<0)$ when $C{\rm{ind,ins}}>0.6$. Hence, for larger cross-correlations ($C>0.6$), there are much more occurrences of positive cross-correlations than negative ones for all the four types of pairs. This striking feature can be attributed to two reasons. The first is that a large proportion of investors react to the same external news in the same direction \cite{Lillo-Moro-Vaglica-Mantegna-2008-NJP}: they buy following good news and sell following bad news. The second is that investors imitate the trading behaviors of others of the same type and rarely imitate other investors of different type. The second reason is rational because the friends of individual (or institutional) investors are more likely individual (or institutional) investors.

\subsection{Eigenvalue spectrum}
\label{subsec:EigSpectrum}

For the correlation matrix $C$ of each stock, we can calculate its eigenvalues, whose density $f_{c}(\lambda)$ is defined as follows \cite{Laloux-Cizean-Bouchaud-Potters-1999-PRL},
\begin{equation}
 f_{c}(\lambda)=\frac{1}{N}\frac{dn(\lambda)}{d\lambda},
 \label{Eq:PDF:Eigenvalues}
\end{equation}
where $n(\lambda)$ is the number of eigenvalues of $C$ less than $\lambda$. If $M$ is a $T\times N$ random matrix with zero mean and unit variance, $f_{c}(\lambda)$ is self-averaging. Particularly, in the limit $N\rightarrow \infty$, $T\rightarrow \infty$ and $Q=T/N\geq1$ fixed, the probability density function $f_{c}(\lambda)$ of eigenvalues $\lambda$ of the random correlation matrix $M$ can be described as \cite{Laloux-Cizean-Bouchaud-Potters-1999-PRL,Plerou-Gopikrishnan-Rosenow-Amaral-Guhr-Stanley-2002-PRE,Sengupa-Mitra-1999-PRE},
\begin{equation}
 f_{c}(\lambda)=\frac{Q}{2\pi\sigma^2}\frac{\sqrt{(\lambda_{\max}-\lambda)(\lambda-\lambda_{\min})}}{\lambda}~,
 \label{Eq:RMTPDF}
\end{equation}
with $\lambda\in[\lambda_{\min}, \lambda_{\max}]$, where $\lambda_{\min}^{\max}$ is given by
\begin{equation}
 \lambda_{\min}^{\max}=\sigma^2(1+1/Q\pm2\sqrt{1/Q})~,
 \label{Eq:RMT:lambda}
\end{equation}
and $\sigma^2$ is equal to the variance of the elements of $M$ \cite{Laloux-Cizean-Bouchaud-Potters-1999-PRL,Sengupa-Mitra-1999-PRE}. The variance $\sigma^2$ is equal to 1 in our normalized data.

\begin{figure}[htb]
  \centering
  \includegraphics[width=8cm]{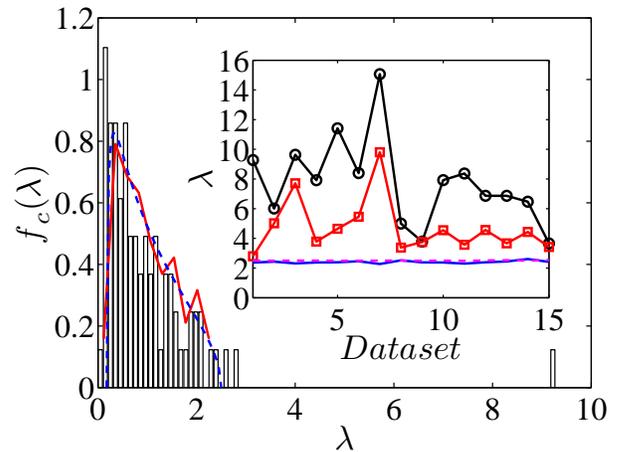}
  \caption{\label{Fig:EigenvalueSpectrum} (Color online) Eigenvalue spectrum of the correlation matrix of inventory variation of investors trading stock 000001 within 1 day time horizon in 2003. The solid line is the spectral density obtained by shuffling independently the buyers and the sellers in such a way to maintain the same number of purchases and sales for each investor as in the real data. The dashed blue line shows the spectral density predicted by the random matrix theory using Eq.~(\ref{Eq:RMTPDF}) with $Q=237/80=2.96$. The inset shows the largest eigenvalue $\lambda_1$ ($\bigcirc$) and the second largest eigenvalue $\lambda_2$ ({\color{red}{$\square$}}) of all 15 investigated data sets from 15 stocks. The solid line indicates the upper thresholds by shuffling experiments, and the dashed line presents the threshold predicted by the random matrix theory.}
\end{figure}

Figure \ref{Fig:EigenvalueSpectrum} illustrates the probability distribution $f_c(\lambda)$ of the correlation matrix of inventory variation of investors trading stock 000001. The solid line is the spectral density obtained by shuffling independently the buyers and the sellers in such a way to maintain the same number of purchases and sales for each investor as in the real data, while the dashed blue line shows the spectral density predicted by the random matrix theory using Eq.~(\ref{Eq:RMTPDF}) with $Q=237/80=2.96$. We find the largest eigenvalue is well outside of the bulk and the second largest eigenvalue also escapes the bulk. The results for other 14 stocks are quite similar. In the inset of Fig.~\ref{Fig:EigenvalueSpectrum}, we plot the largest eigenvalues $\lambda_1$ and the second largest eigenvalues $\lambda_2$ for all the 15 stocks. We find that all the largest eigenvalues are well above the upper thresholds determined from shuffling experiments and the thresholds $\lambda_{\max}$ in Eq.~(\ref{Eq:RMT:lambda}) predicted by the random matrix theory. Moreover, all the second largest eigenvalues are above the two threshold lines and some of them are well above the thresholds. These findings indicates that both the largest and the second largest eigenvalues carry information about the investors, which is different from different from the results of the Spanish stock market, where only the largest eigenvalue is larger than the up thresholds while the second largest eigenvalue is within the bulk \cite{Lillo-Moro-Vaglica-Mantegna-2008-NJP}. This discrepancy can be attributed to the difference of the two markets and the fact that our analysis contains both individuals and institutions while Lillo et al studies only firms.

\subsection{Distribution of eigenvector components}
\label{S2:infor}

If there is no information contained in an eigenvalue, the normalized components of its associated eigenvector should conform to a Gaussian distribution \cite{Laloux-Cizean-Bouchaud-Potters-1999-PRL,Plerou-Gopikrishnan-Rosenow-Amaral-Stanley-1999-PRL,Plerou-Gopikrishnan-Rosenow-Amaral-Guhr-Stanley-2002-PRE}:
\begin{equation}
 f(u) = \frac{1}{\sqrt{2\pi}}\exp\left(-\frac{u^2}{2}\right).
 \label{Eq:PDF:vector:Gaussian}
\end{equation}
Since the empirical eigenvalue distribution $f_c(\lambda)$ deviates from the theoretic expression (\ref{Eq:PDF:Eigenvalues}) from the random matrix theory with two large eigenvalues outside the bulk of the distribution, it is expected that the associated eigenvectors also contain certain information.

\begin{figure}[htb]
\centering
\includegraphics[width=8cm]{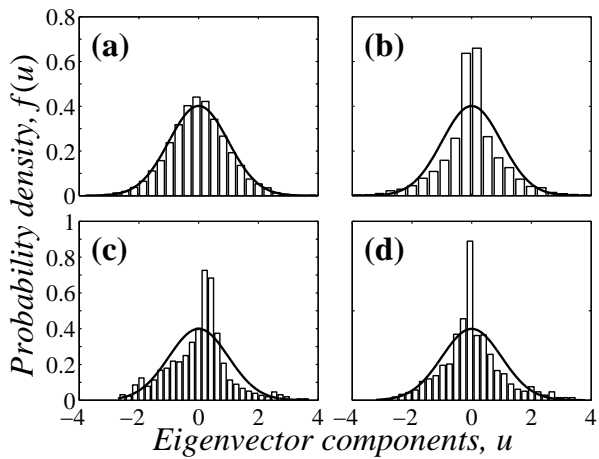}
\caption{\label{Fig:PDF:eigenvector} Distribution of eigenvector components $u$: (a) all the eigenvectors associated with the eigenvalues in the bulk $\lambda_{\min}<\lambda<\lambda_{\max}$ after normalization for each eigenvector for stock 000001, (b) same as (a) but for stock 200625, (c) all the eigenvectors associated with the largest eigenvalues $\lambda_1$ after normalization for all the stocks, and (d) all the eigenvectors associated with the second largest eigenvalues $\lambda_2$ after normalization for all the stocks. The solid lines show the Gaussian distribution predicted by the random matrix theory.}
\end{figure}

For correlation matrices of financial returns, the components of an eigenvector with the eigenvalue $\lambda$ in the bulk of its distribution ($\lambda_{\min}<\lambda<\lambda_{\max}$) are distributed according to Eq.~(\ref{Eq:PDF:vector:Gaussian}) \cite{Laloux-Cizean-Bouchaud-Potters-1999-PRL,Plerou-Gopikrishnan-Rosenow-Amaral-Stanley-1999-PRL,Plerou-Gopikrishnan-Rosenow-Amaral-Guhr-Stanley-2002-PRE}. Panels (a) and (b) in Fig.~\ref{Fig:PDF:eigenvector}show the empirical distributions of the eigenvector components $u$ with the eigenvalues in the bulk for two typical stocks. Rather than analyzing the vector for one eigenvector, we normalized the components of each eigenvector and put all the eigenvectors together to gain better statistics, since each eigenvector has only 80 components. We find that the distributions of 10 stocks are well consistent with the Gaussian, while other 5 stocks exhibit high peaks in the center. 

For deviating eigenvalues $\lambda_1$ and $\lambda_2$, the distribution for each stock is very noisy and deviates from Gaussian. 
We treat the components of the 15 eigenvectors as a sample to have better statistics. The two distributions obtained are illustrated in Fig.~\ref{Fig:PDF:eigenvector}(c) and (d). It is evident that both deviate from the Gaussian distribution and the distribution for $\lambda_1$ is more skewed.

\subsection{Information in eigenvectors for deviating eigenvalues}
\label{S2:info}

We have shown that the largest and the second largest eigenvalues deviate from the RMT prediction and the distributions of their eigenvector components are not Gaussian. It implies that these eigenvectors carry some information. For $u(\lambda_2)$, it is not clear what kind of information they carry. We find no evident dependence of the magnitude of $u_i(\lambda_2)$ on the average absolution inventory variation $\langle{|v_i|}\rangle$, the total variation $\sum{v_i}$, or the maximal absolute variation $\max\{|v_i|\}$. Same conclusion is obtained for $u_i(\lambda_1)$, which differs from the conclusion that the eigenvector components of the return correlation matrix depend on the market capitalization in a logarithmic form \cite{Plerou-Gopikrishnan-Rosenow-Amaral-Guhr-Stanley-2002-PRE}. In addition, as we will show in the next section that the investors can be categorized into three trading types. We also find no relation between the trading strategy category and the magnitude of the vector component $u(\lambda_2)$. We thus focus on extracting the information from $u(\lambda_1)$.

\begin{figure}[htb]
  \centering
  \includegraphics[width=8cm]{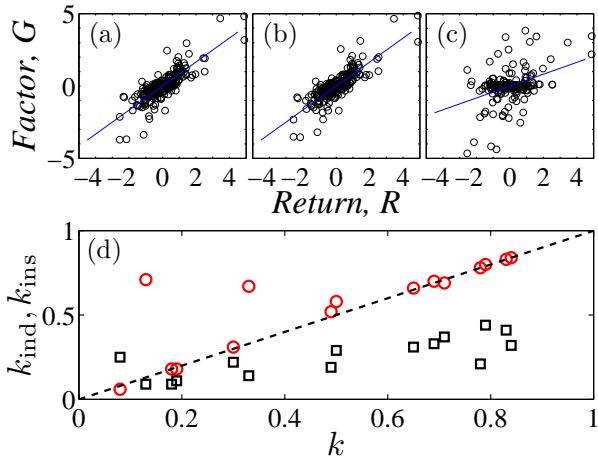}
  \caption{\label{Fig:R:G} (Color online) Influence of the most active investors trading stock 000001 on the price fluctuations for all the investigated investors (a), for individual investors (b), and for institutional investors (c), where the slopes are $k=0.83\pm0.04$, $k_{\rm{ind}}= 0.83\pm0.04$, and $k_{\rm{ins}}= 0.41\pm 0.06$, respectively. Panel (d) plots $k_{\rm{ind}}$ and $k_{\rm{ins}}$ against $k$, where each symbol corresponds to a stock.}
\end{figure}

For the correlation matrix whose elements are the correlation coefficients of price fluctuations of two stocks, the eigenvector of the largest eigenvalue contains market information \cite{Laloux-Cizean-Bouchaud-Potters-1999-PRL,Plerou-Gopikrishnan-Rosenow-Amaral-Stanley-1999-PRL}. The market information indicates the collective behavior of stock price movements \cite{Plerou-Gopikrishnan-Rosenow-Amaral-Stanley-2001-PA}, which can be unveiled by the projection of the time series on the eigenvector \cite{Plerou-Gopikrishnan-Rosenow-Amaral-Guhr-Stanley-2002-PRE}. We follow this approach and calculate the projection $G(t)$ of the time series $V_{i}(t)=[{v_{i}(t)-\langle v_{i}(t)\rangle}]/{\sigma_{i}}$ on the eigenvector $u(\lambda_1)$ corresponding to the first eigenvalue \cite{Lillo-Moro-Vaglica-Mantegna-2008-NJP}:
\begin{equation}
 G(t) = \sum_i V_i(t)\times u_i(\lambda_1)(t),
 \label{Eq:factor}
\end{equation}
The projection $G$ can be called the factor associated with the largest eigenvalue \cite{Lillo-Moro-Vaglica-Mantegna-2008-NJP}. We plot the factor $G(t)$ against the normalized return $R(t)$ for stock 000001 in Fig.~\ref{Fig:R:G}(a). There is a nice linear dependence between the two variables and a linear regression gives the slope $k=0.83\pm0.04$. It indicates that these most active investors have dominating influence on the price fluctuations.

Panels (b) and (c) of Fig.~\ref{Fig:R:G} illustrate the relation between the factor and the return for individuals and institutions. Linear regression gives $k_{\rm{ind}}= 0.83\pm0.04$ and $k_{\rm{ins}}= 0.41\pm 0.06$. Comparing (b) and (c) with (a), we find that the influence of individuals matches excellently with the whole sample, which can be quantified by the facts that $k_{\rm{ind}}=k$ and $k_{\rm{ind}}<k$. The results are similar for other stocks. 
The resulting $k_{\rm{ind}}$ and $k_{\rm{ins}}$ are plotted in Fgi.~\ref{Fig:R:G} against $k$ for all the 15 stocks. For individual investors, we find that $k_{\rm{ind}}=k$ for 13 stocks and $k_{\rm{ind}}>k$ for 2 stocks. In contrast, we find that $k_{\rm{ins}}<k$ except for one stock.

\section{Inventory variation and stock return}
\label{S1:CVR}

\subsection{Categorization of investors}
\label{S2:categorization}

\begin{figure*}[htb]
\centering
\includegraphics[width=5.5cm]{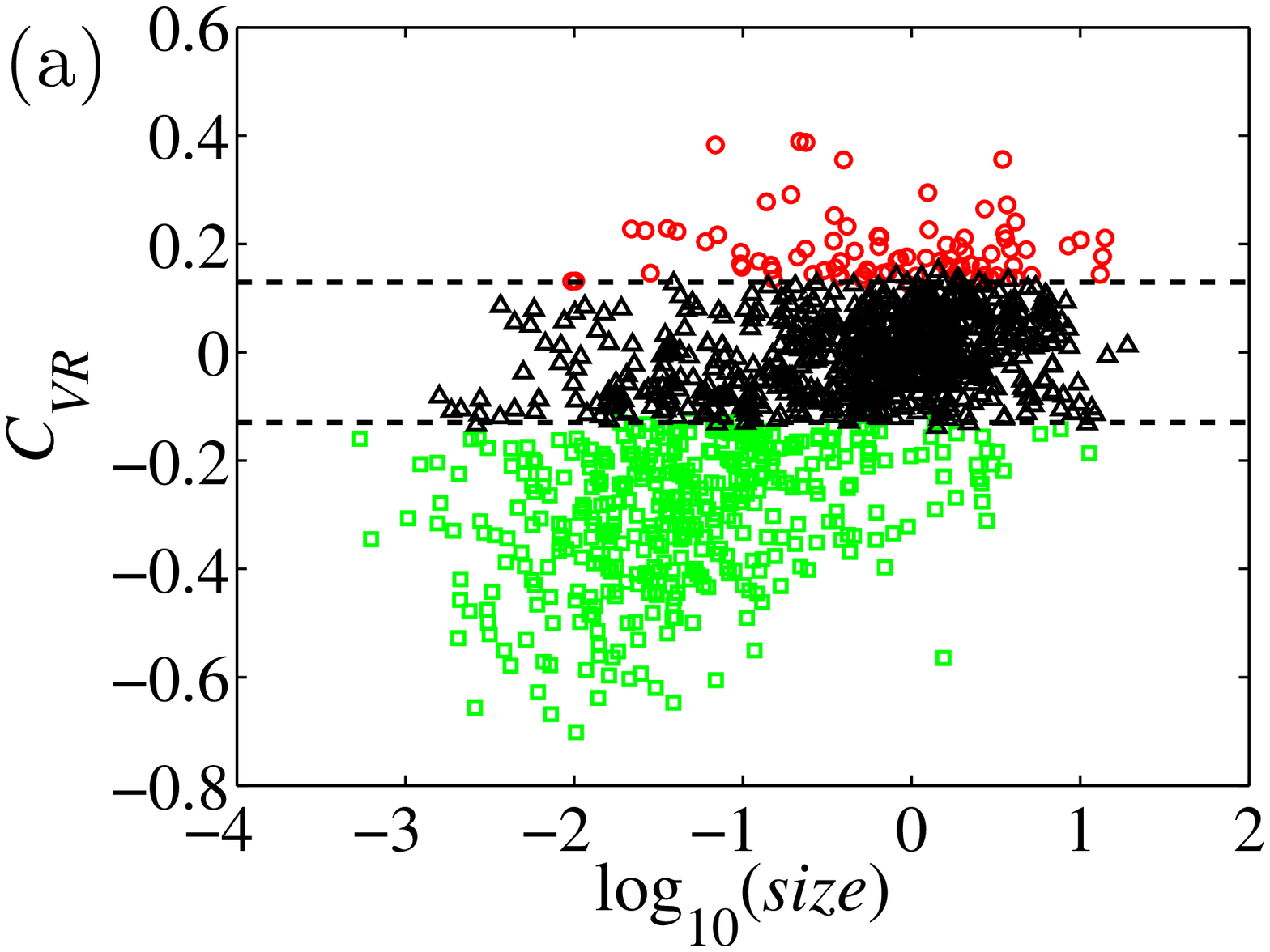}
\includegraphics[width=5.5cm]{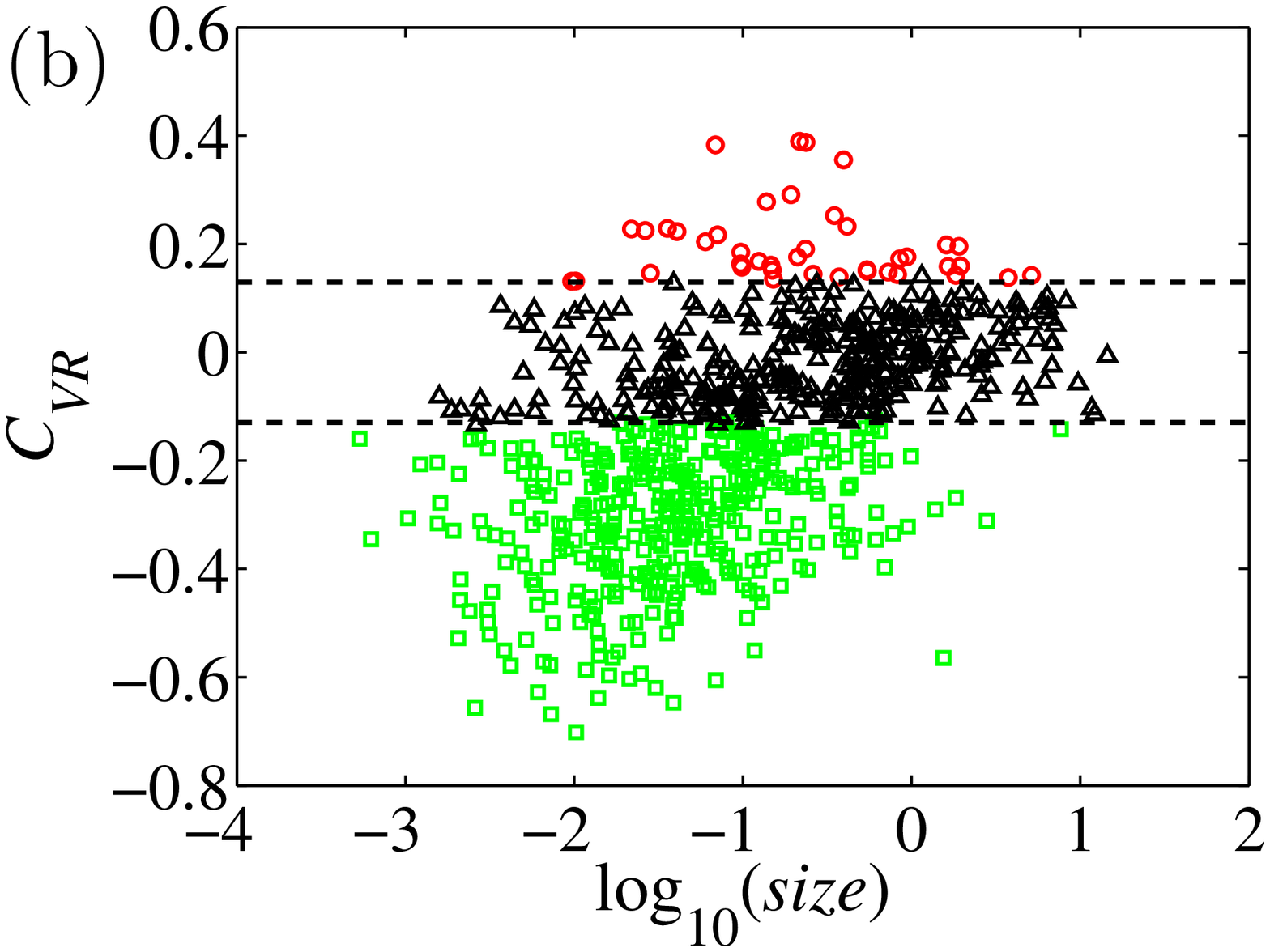}
\includegraphics[width=5.5cm]{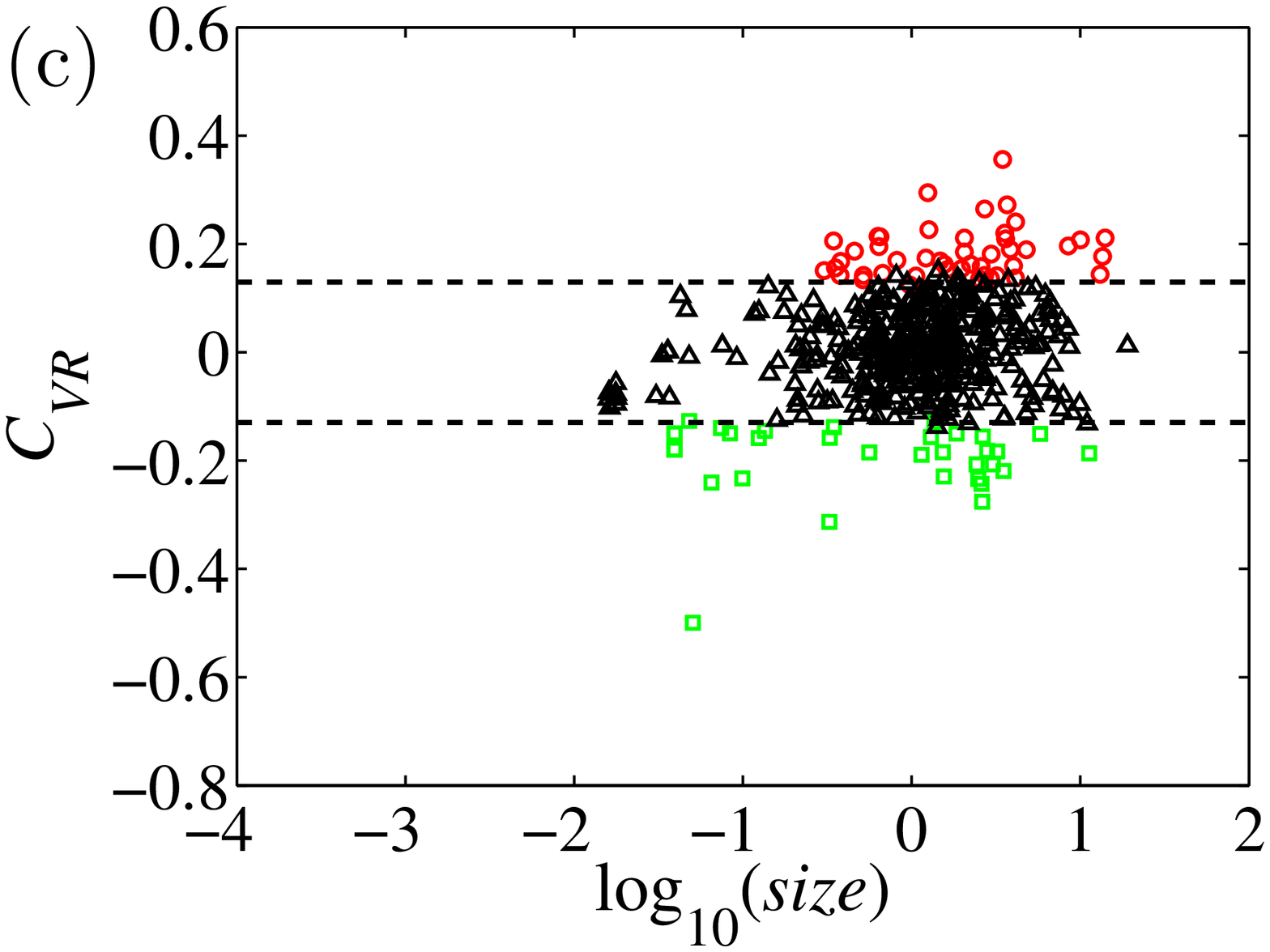}
\includegraphics[width=5.5cm]{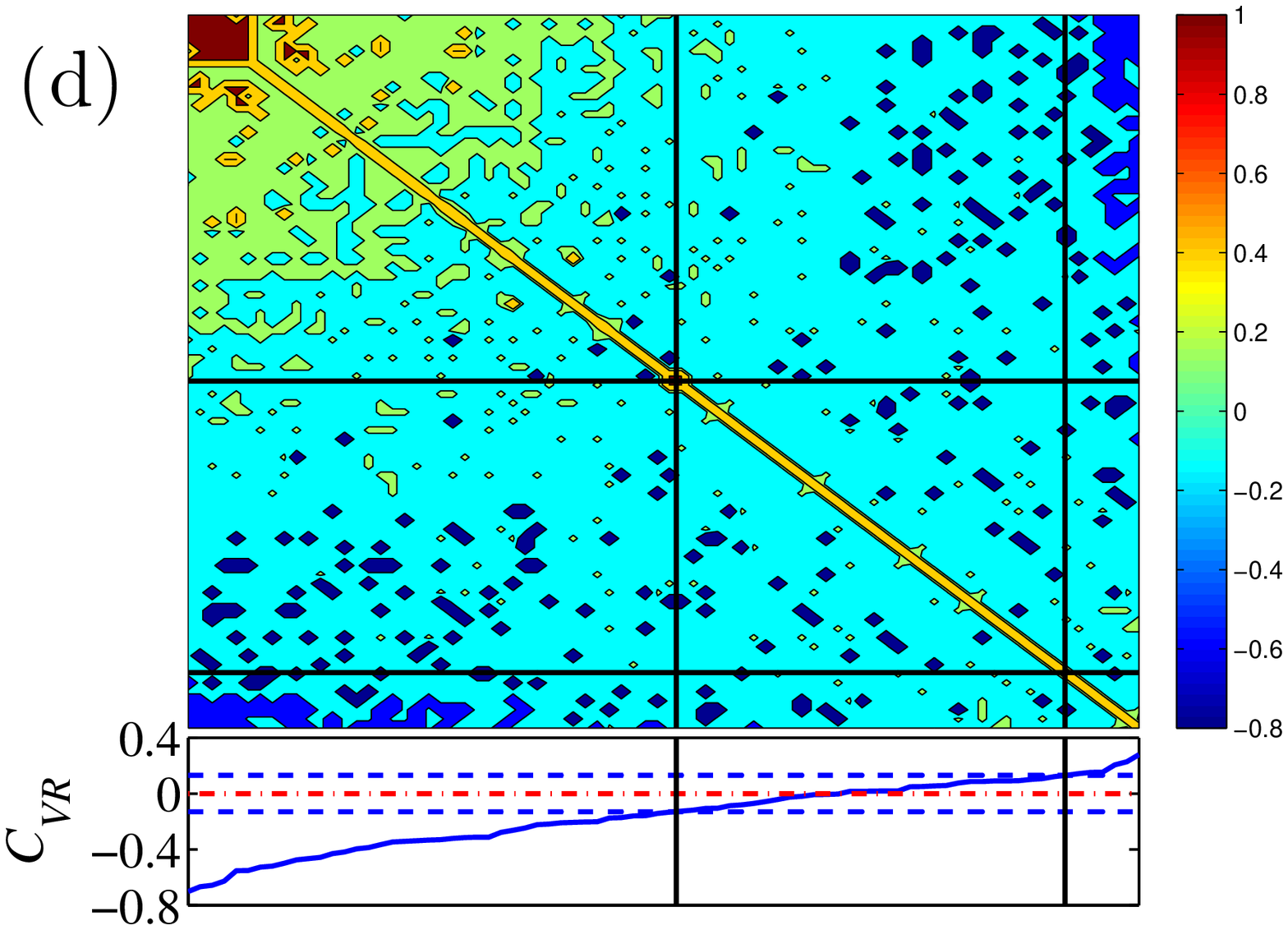}
\includegraphics[width=5.5cm]{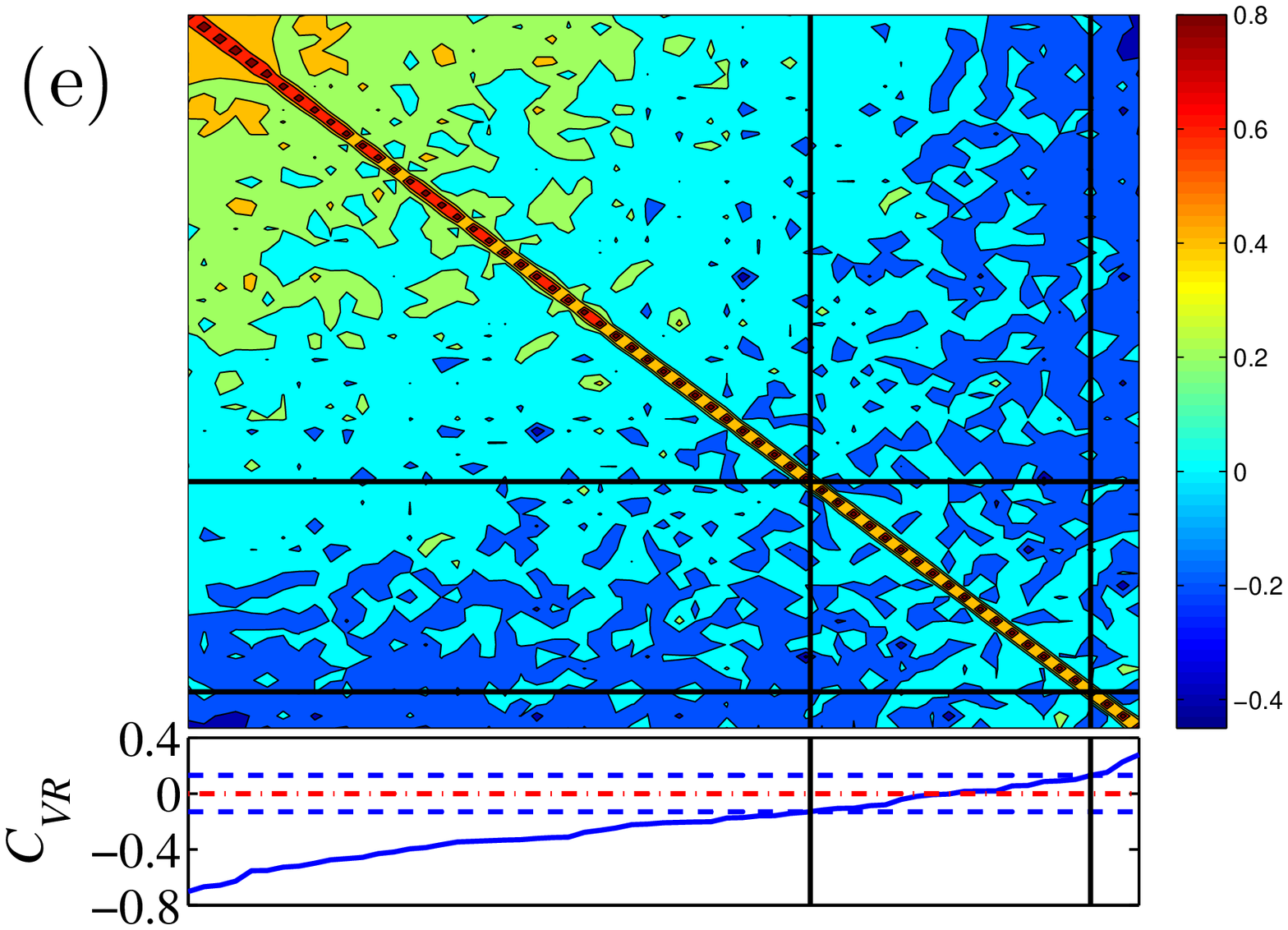}
\includegraphics[width=5.5cm]{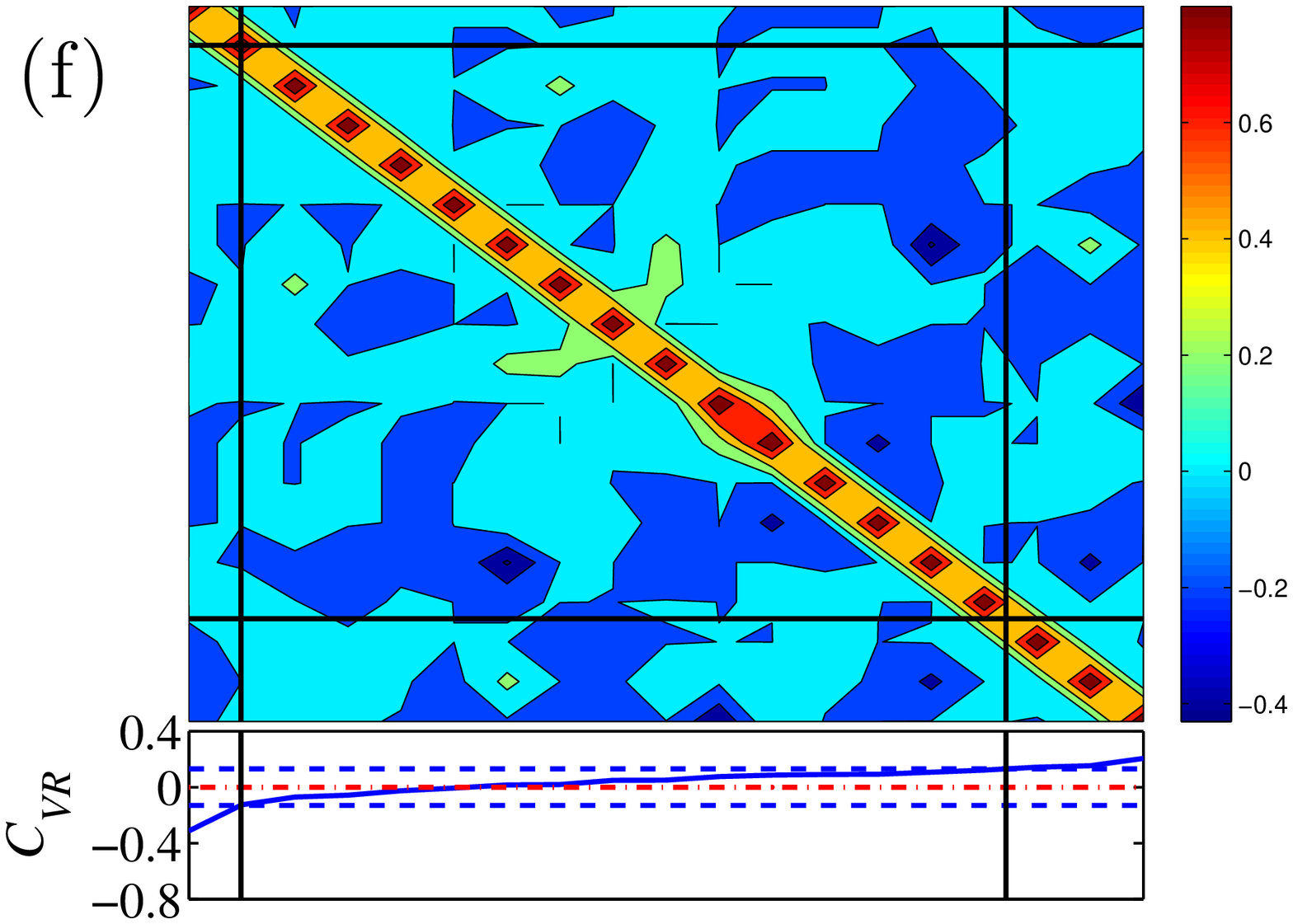}
 \caption{\label{Fig:Categorization:test} (Color online) Panels (a-c) show the scatter plots of $C_{VR}$ versus a proxy of the size of the investor. For each stock, the proxy is the ratio of the value exchanged by the investor (scaled by a factor $10^4$) to the capitalization of the stock. Each marker refers to an investor trading a specific stock. The three kinds of markers refer to investors whose inventory variations are positively correlated ({\color{red}{$\bigcirc$}}), negatively correlated ({\color{green}{$\square$}}), or uncorrelated ($\vartriangle$) with returns according to the block bootstrap analysis. The two dashed lines indicate the $2\sigma$ threshold calculated using Eq.~(\ref{Eq:threshold}). Panels (d-f) are contour plots of the correlation matrix of daily inventory variation of investors trading the stock 000001. We have sorted the investors into rows and columns according to their cross-correlation coefficients of inventory variation with its price return $C_{VR}$. The evolution of $C_{VR}$ in the same order as in the matrix is shown in the bottom panel, where the dashed lines bound the $\pm 2\sigma$ significance intervals.}
\end{figure*}

Following Ref.~\cite{Lillo-Moro-Vaglica-Mantegna-2008-NJP}, we divide the investors into three categories according to the cross-correlation coefficient $C_{V_{i}R}$ between the inventory variation $V_{i}$ and the stock return $R$. The investor $i$ belongs to the trending or reversing category if its inventory variation is positively or negatively correlated with the return. We use a wieldy significant threshold to categorize the investors:
\begin{equation}
 \pm2\sigma=\pm2/\sqrt{N_T}~,
 \label{Eq:threshold}
\end{equation}
where $N_{T}$ is the number of time records for each time series~\cite{Lillo-Moro-Vaglica-Mantegna-2008-NJP}. We also verify the robustness of Eq.~(\ref{Eq:threshold}) by comparing the experimental results with the results of a null hypothesis based on a block bootstrap of both $R$ and $V$. In this regard, 1000 block bootstrap replicas with a block length of 20 are performed. For each investor, we have checked whether the estimated correlation with return exceeds the 0.97725 quantile or is smaller than the 0.02275 quantile of the correlation distribution obtained from bootstrap replicas. The results are shown in Fig.~\ref{Fig:Categorization:test}(a-c). There are 1211 investors in the whole sample in Fig.~\ref{Fig:Categorization:test}(a), including 453 institutional investors in Fig.~\ref{Fig:Categorization:test}(b) and 758 individual investors in Fig.~\ref{Fig:Categorization:test}(c).

As shown in the last row of Table \ref{TB:BasicStat}, the numbers of the three kinds of investors (trending, reversing and uncategorized) are 46, 34 and 373 for institutional investors and 41, 381 and 336 for individual investors, respectively. We find that most institutional investors are uncategorized and there are more trending investors than reversing investors. These results are different from the Spanish case, where only one-third investors are uncategorized and the number of reversing firm investors is about three times the number of trending firm investors \cite{Lillo-Moro-Vaglica-Mantegna-2008-NJP}. In contrast, about half individuals are reversing investors and only 6\% individuals are trending investors. The observation that most investors are uncategorized is probably due to the fact that the Chinese market was emerging and its investors are not experienced. Comparing individuals and institutions, we find a larger proportion of individuals exhibiting a reversing behavior. It indicates that these individuals buy when the price drops and sell when the price rises in the same day. This finding is very interesting since it explains the worse performance in stock markets \cite{Barber-Odean-2000-JF,Barber-Lee-Liu-Odean-2009-RFS}.

The empirical evidence for the significant cross-correlation between inventory variation $V_i(t)$ of trending and reversing investors and stock return $R(t)$ leads us to adopt a linear model for the dynamics of inventory variation as a first approximation \cite{Lillo-Moro-Vaglica-Mantegna-2008-NJP}:
\begin{equation}
 V_i(t) = \gamma_iR(t) + \epsilon_i,
 \label{Eq:Model:v:R}
\end{equation}
where $\gamma_i$ is proportional to the cross-correlation coefficient $C_{V_i,R}$. It follows immediately that the cross-correlation coefficient between the inventory variations of two investors are
\begin{equation}
 C_{ij}=C_{V_i,V_i} = \gamma_i\gamma_j.
 \label{Eq:Model:Cij:gamma:ij}
\end{equation}
If two investors belong to the same category, either trending ($\gamma_i>2\sigma$ and $\gamma_j>2\sigma$ significantly) or reversing ($\gamma_i<-2\sigma$ and $\gamma_j<-2\sigma$ significantly), the value of $C_{ij}$ is expected to be significantly positive. On the contrary, if two investors belong respectively to the trending and reversing categories, the value of $C_{ij}$ is expected to be significantly negative. To show the performance of the model, we plot the contours of the correlation matrix of inventory variation for all investors, for individuals and for institutions, where the investors are sorted according to their cross-correlation coefficients $C_{VR}$ of the inventory variation with the price return. Figure \ref{Fig:Categorization:test}(d) shows that the left-top corner gives large positive $C_{ij}$ values and the left-bottom and right-top corners gives large negative $C_{ij}$ values, as expected. Figure~\ref{Fig:Categorization:test}(e) give better results for individual investors, validating the linear model (\ref{Eq:Model:v:R}). The results in Fig.~\ref{Fig:Categorization:test}(f) are worse for institutional investors, which is due to the fact that most $C_{VR}$ values are small for institutions, as illustrated in Fig.~\ref{Fig:Categorization:test}(c). However, Fig.~\ref{Fig:Categorization:test}(f) doses not invalidate the linear model, since there are only three trending institutions and one reversing institution. 
Indeed, the situation is quite similar for other individual stocks with very few investors of the same category, as shown in Table \ref{TB:BasicStat}.

\subsection{Causality}
\label{S2:causality}

In Sec. \ref{S2:categorization}, we have shown that the inventory variation $V_i(t)$ and the stock return $R(t)$ have significant positive or negative correlation for part of the investors. It is interesting to investigate the lead-lag structure between these two variables. For the largest majority of reversing and trending firms in the Spanish stock market, it is found that returns Granger cause inventory variation but not vice versa at the day or intraday level, and the Granger causality disappears over longer time intervals \cite{Lillo-Moro-Vaglica-Mantegna-2008-NJP}. Here, we aim to study the same topic for both individual and institutional investors.

\begin{figure*}[htb]
\centering
\includegraphics[width=5.5cm]{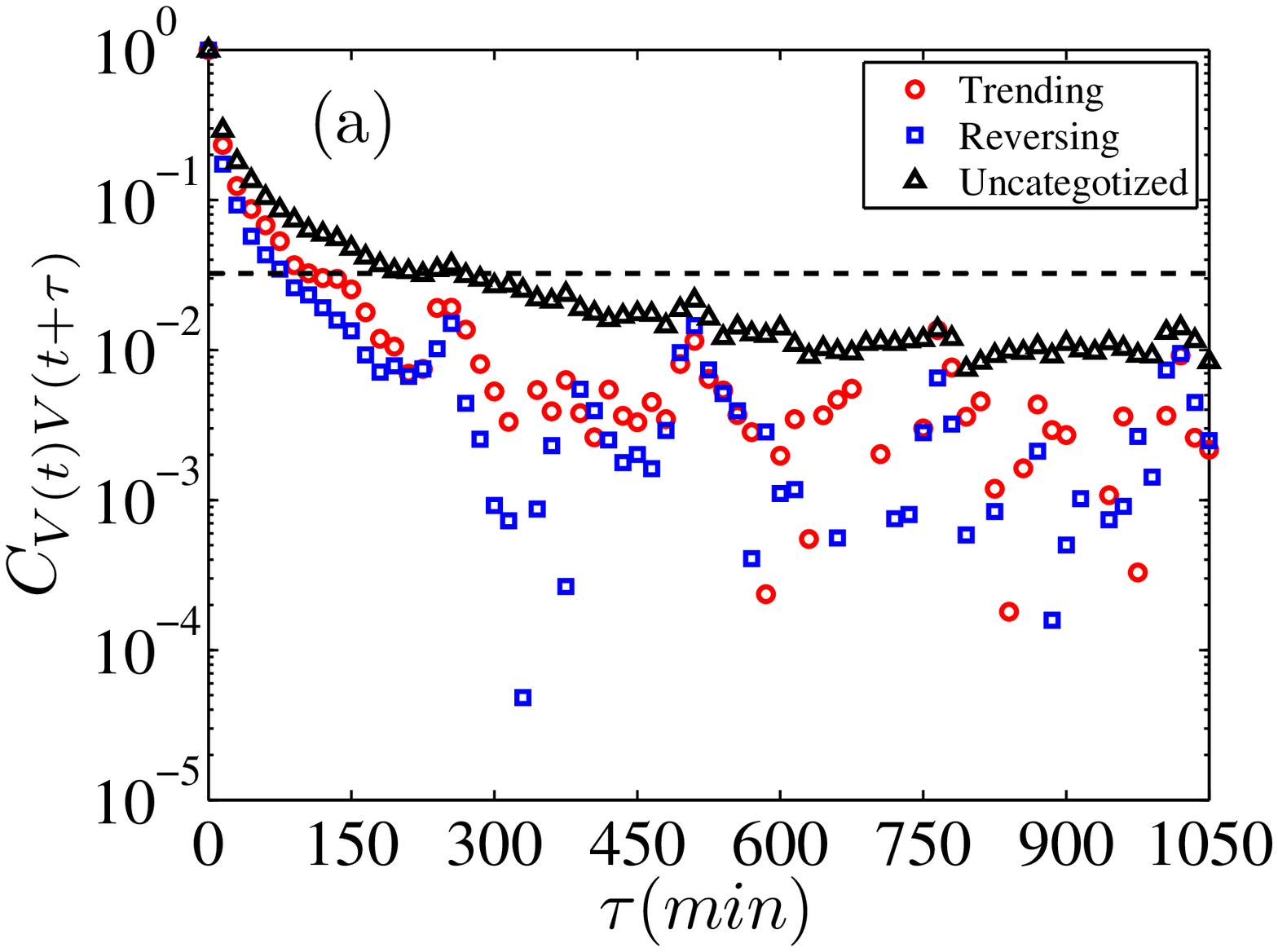}
\includegraphics[width=5.5cm]{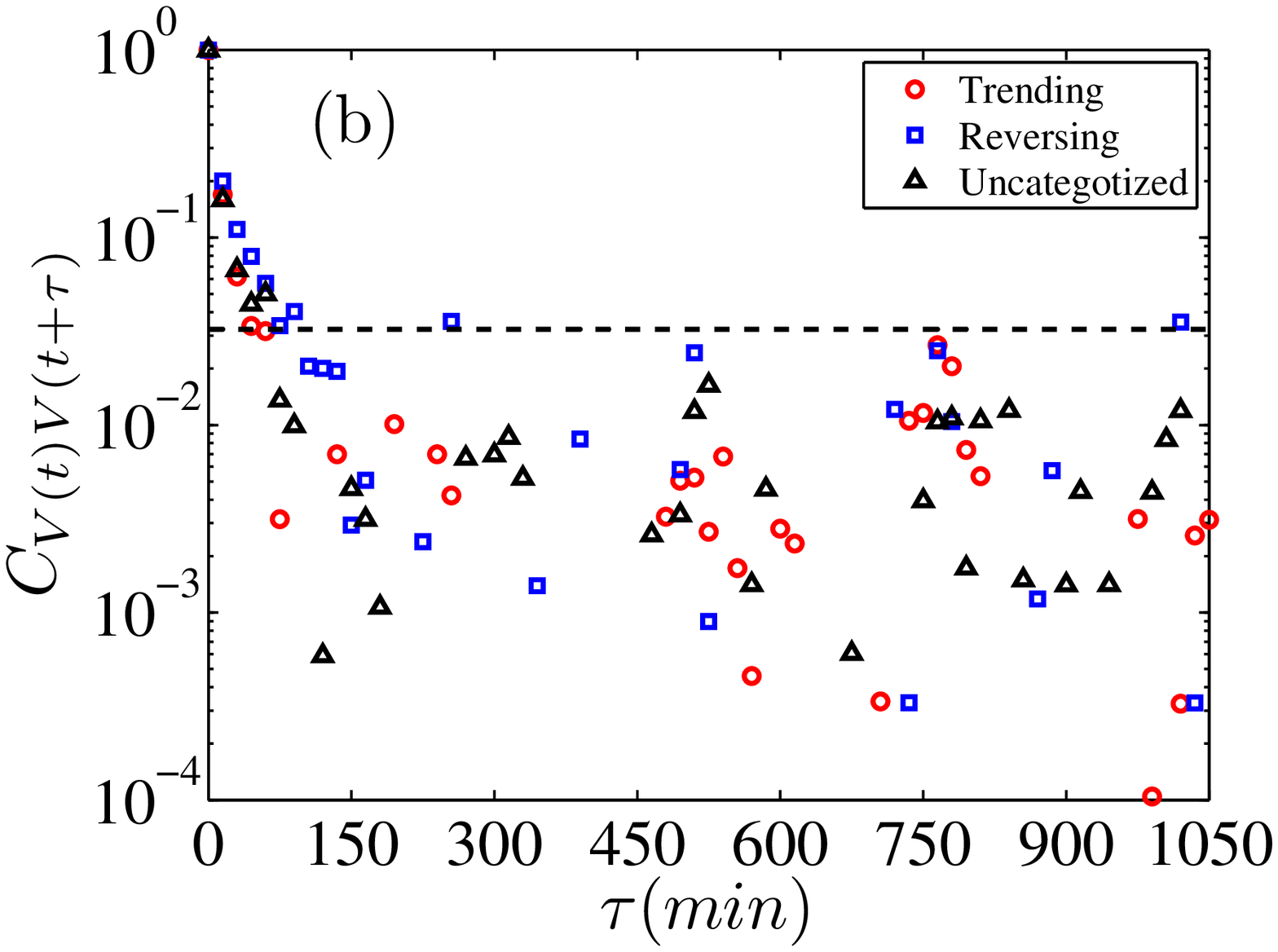}
\includegraphics[width=5.5cm]{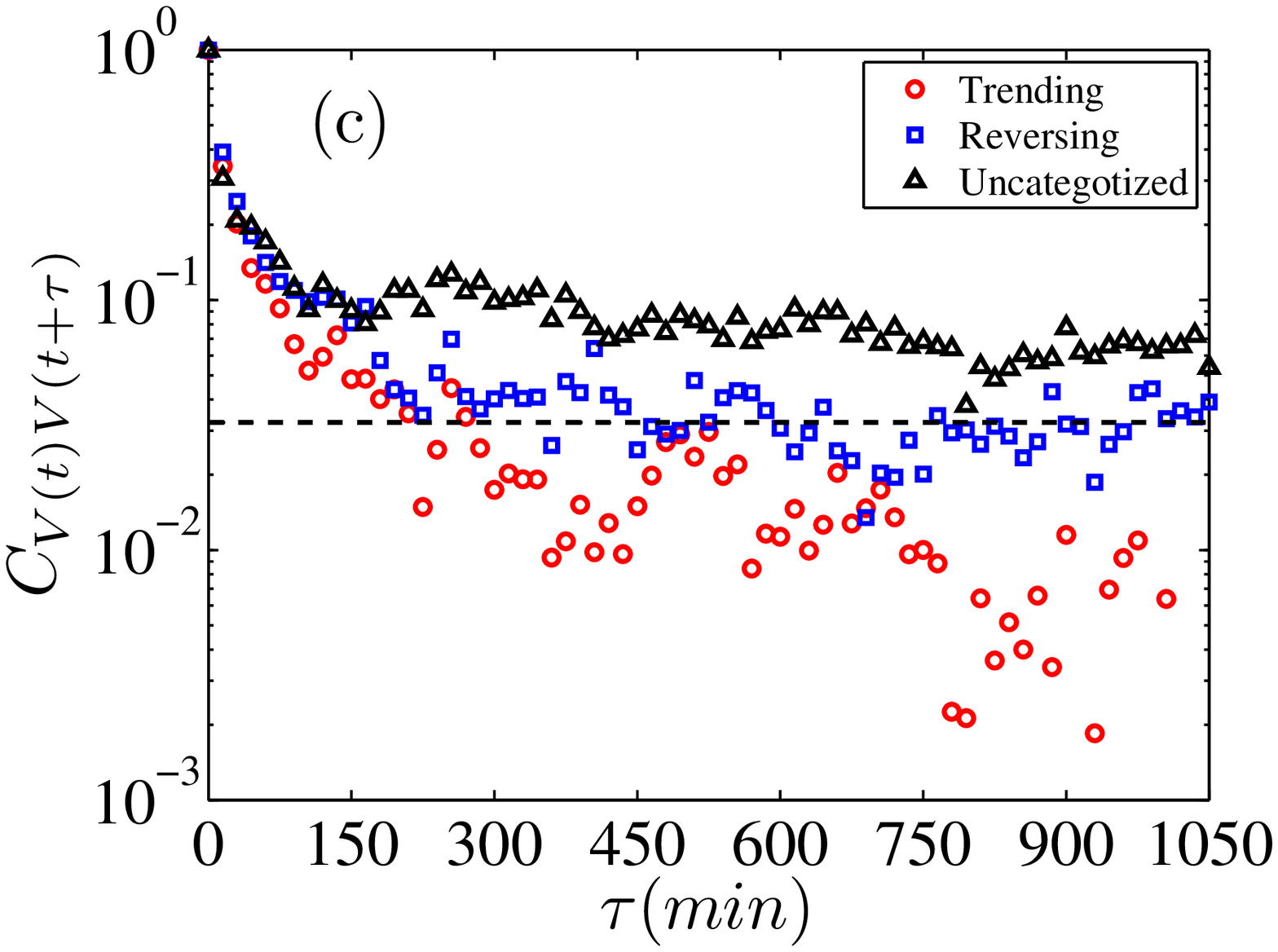}
\includegraphics[width=5.5cm]{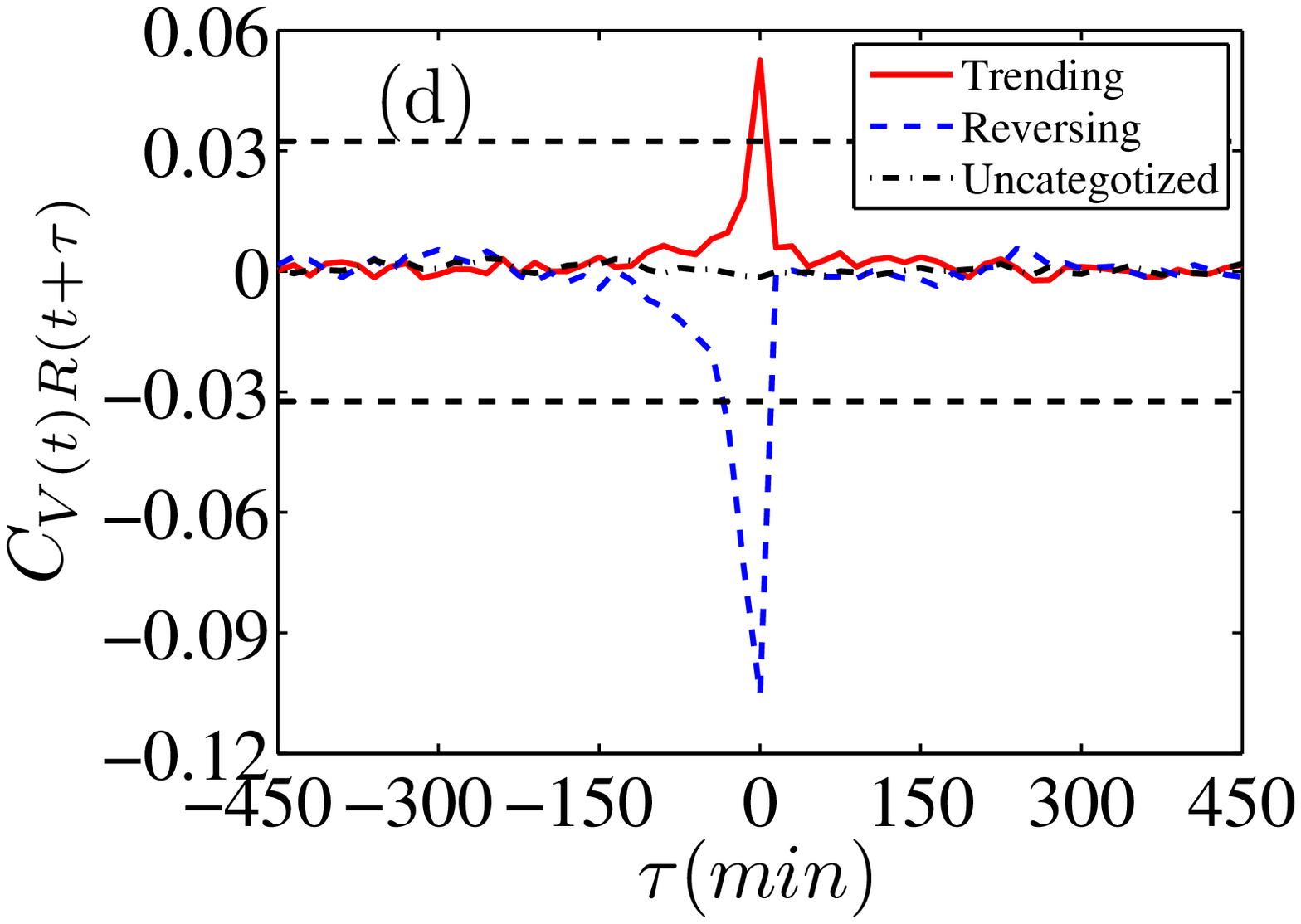}
\includegraphics[width=5.5cm]{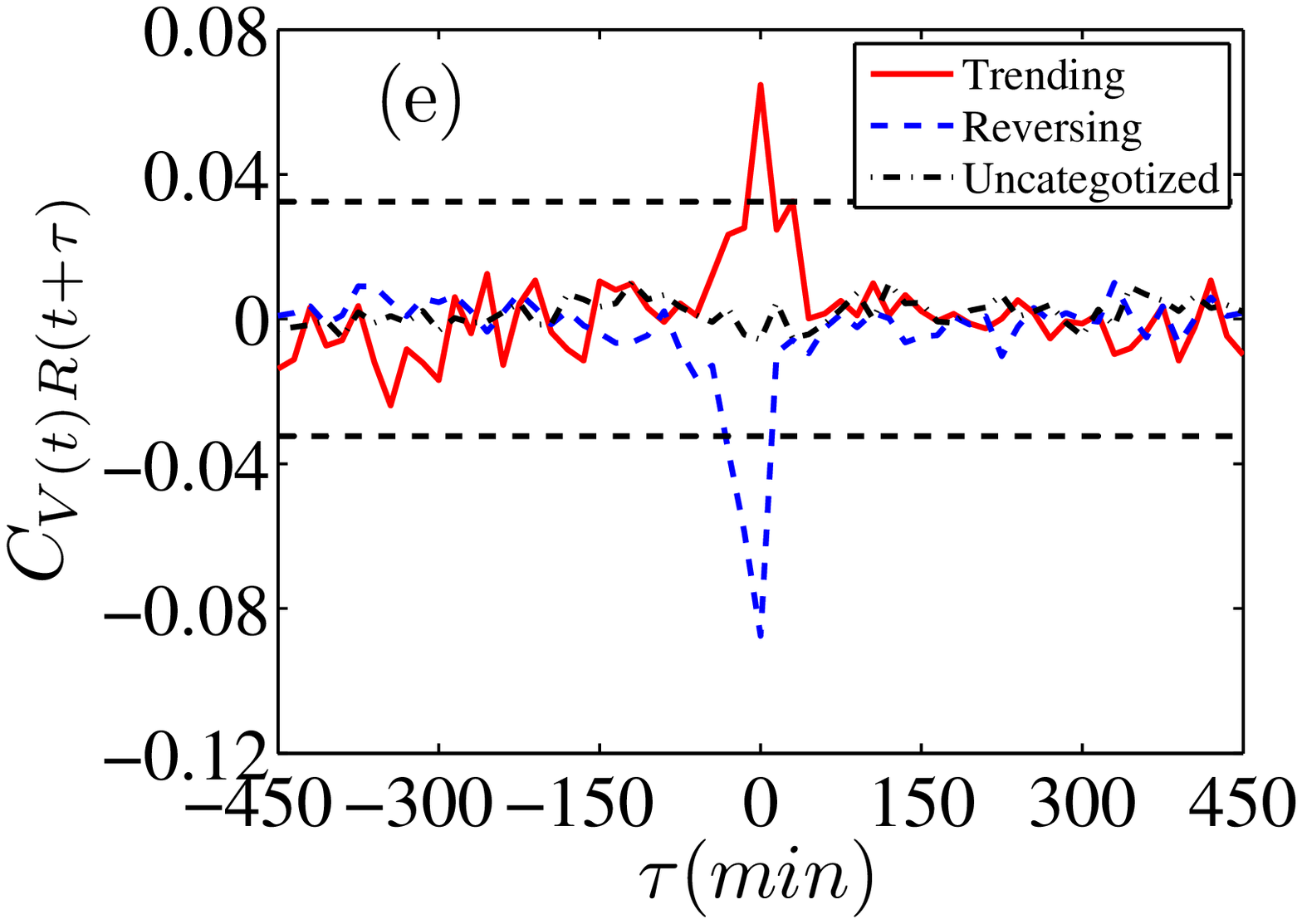}
\includegraphics[width=5.5cm]{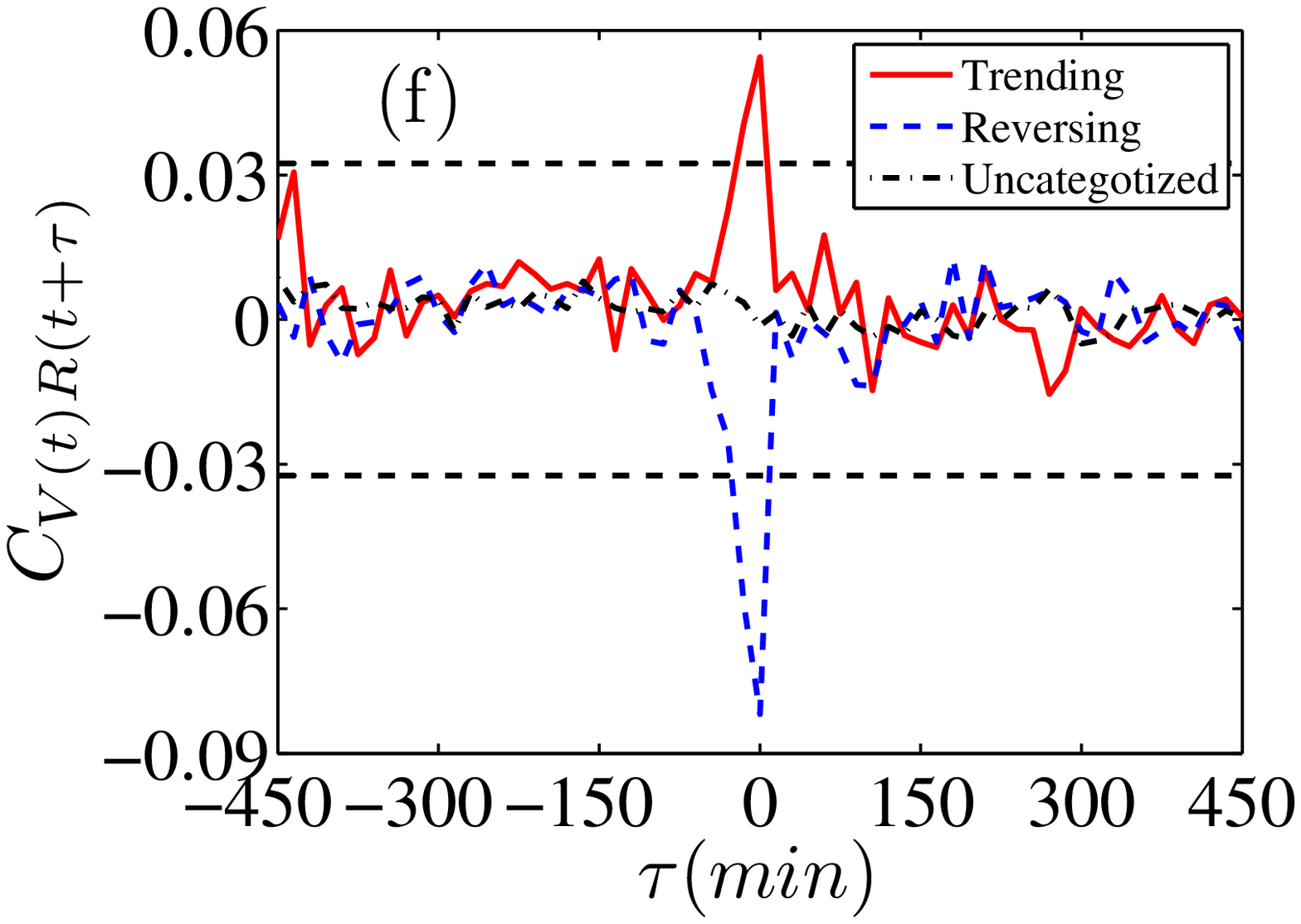}
\includegraphics[width=5.5cm]{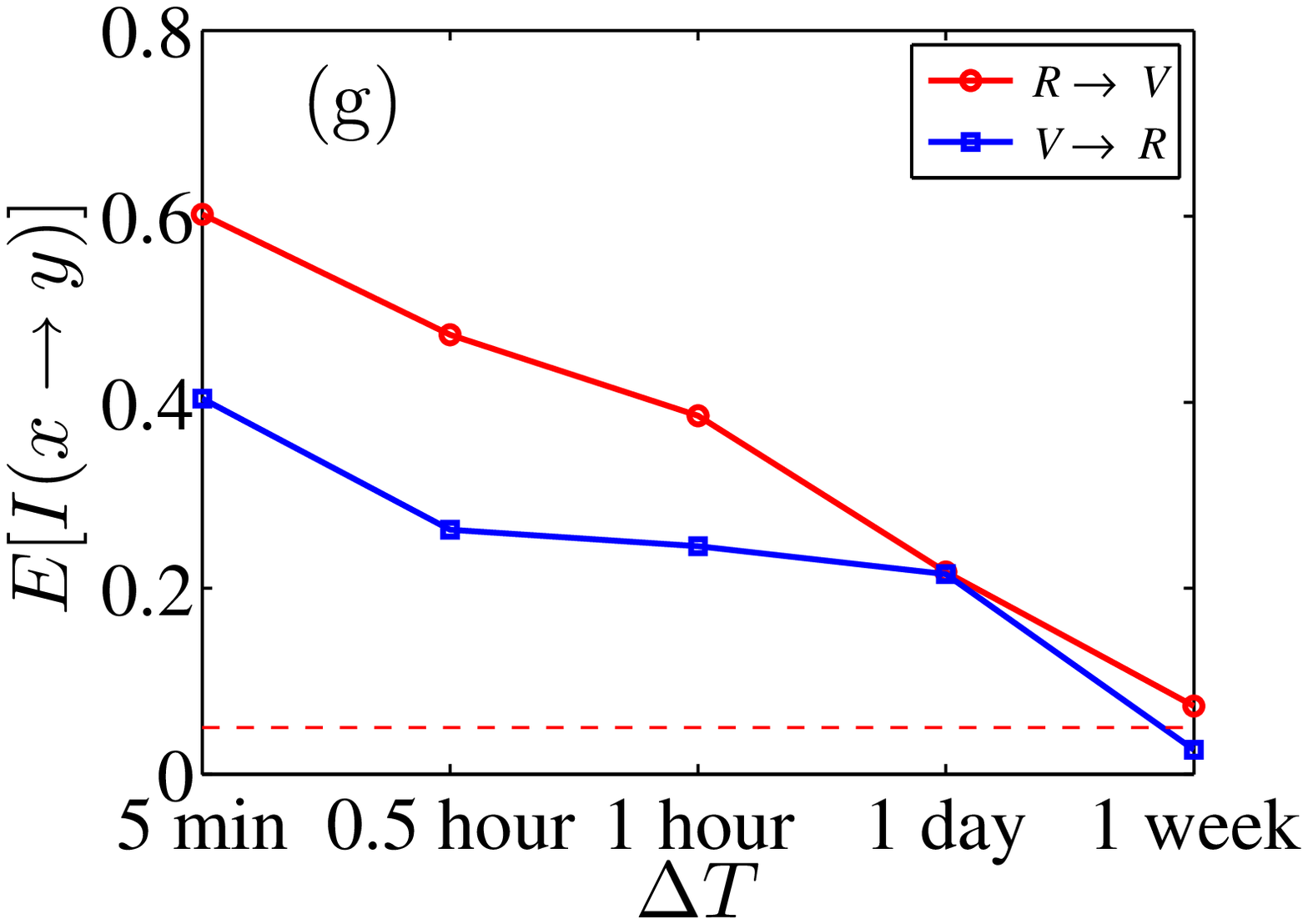}
\includegraphics[width=5.5cm]{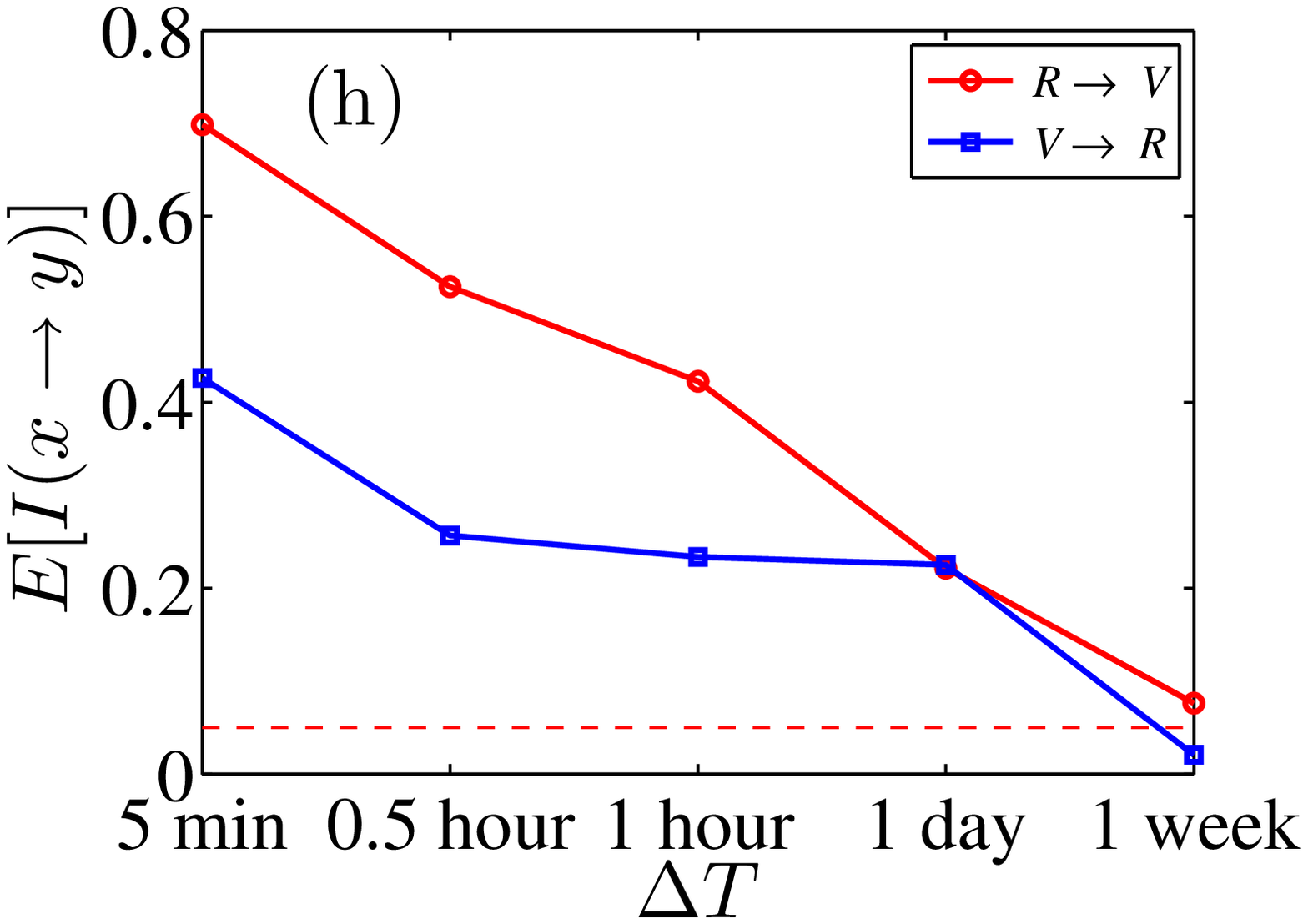}
\includegraphics[width=5.5cm]{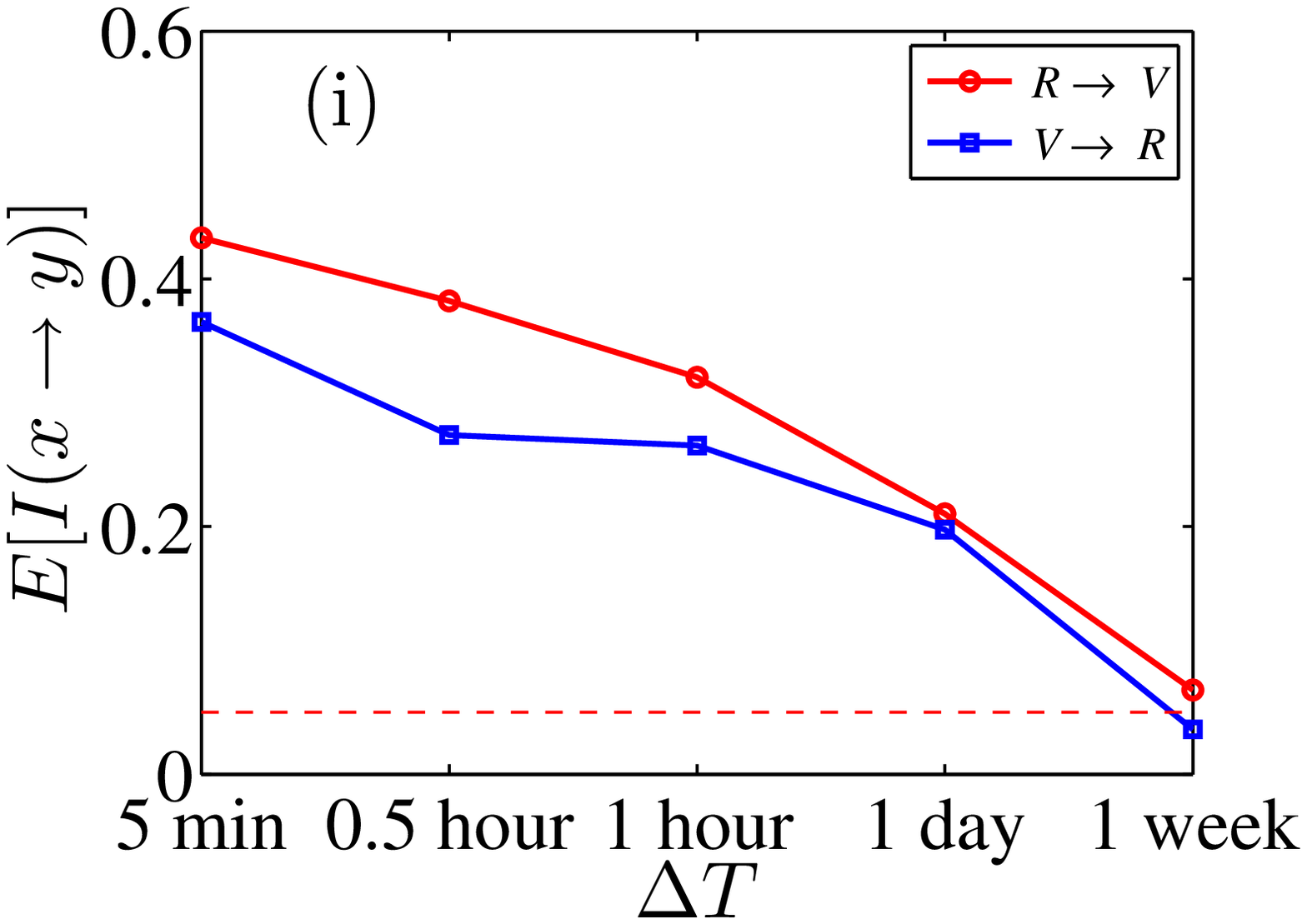}
\includegraphics[width=5.5cm]{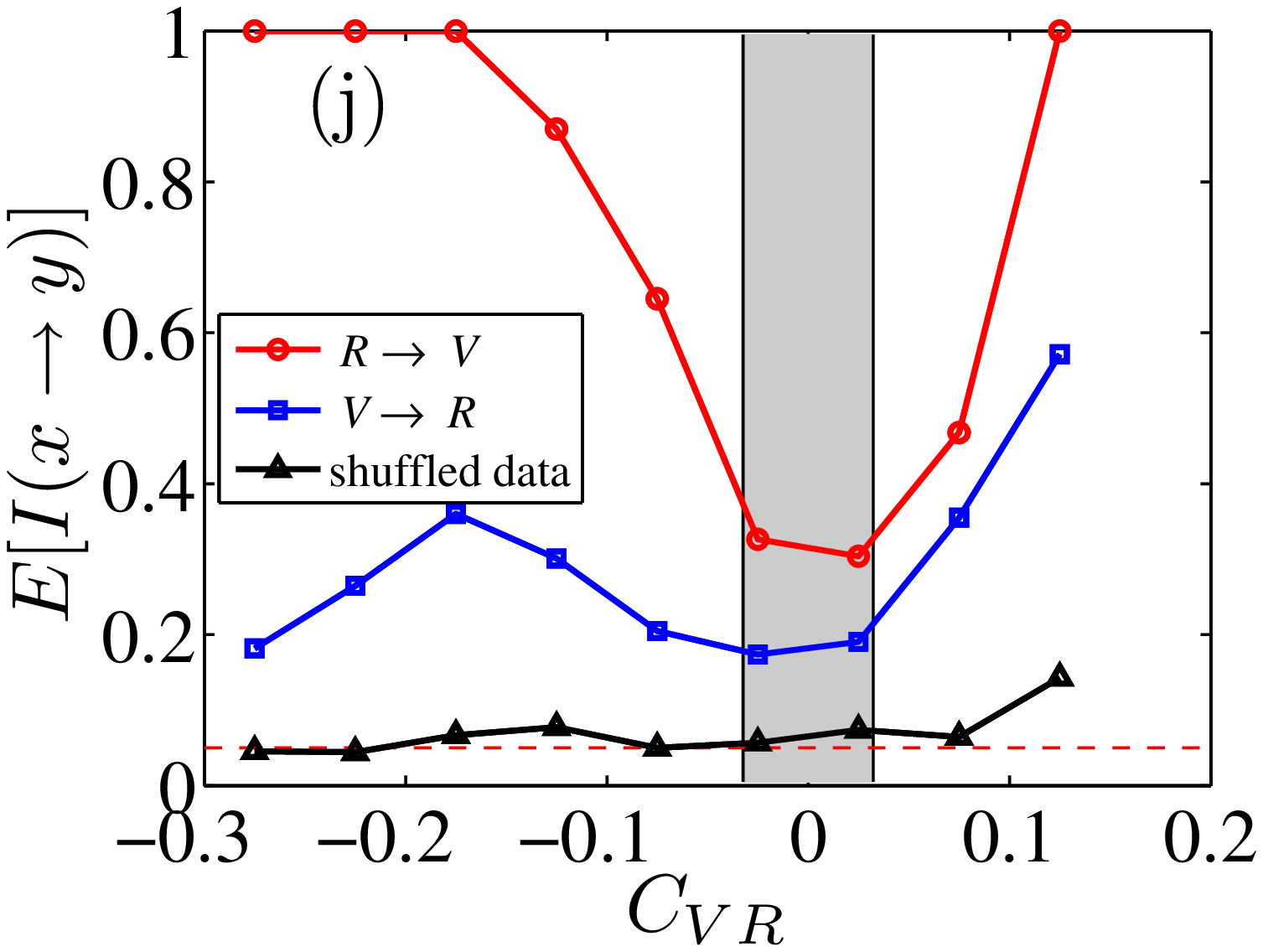}
\includegraphics[width=5.5cm]{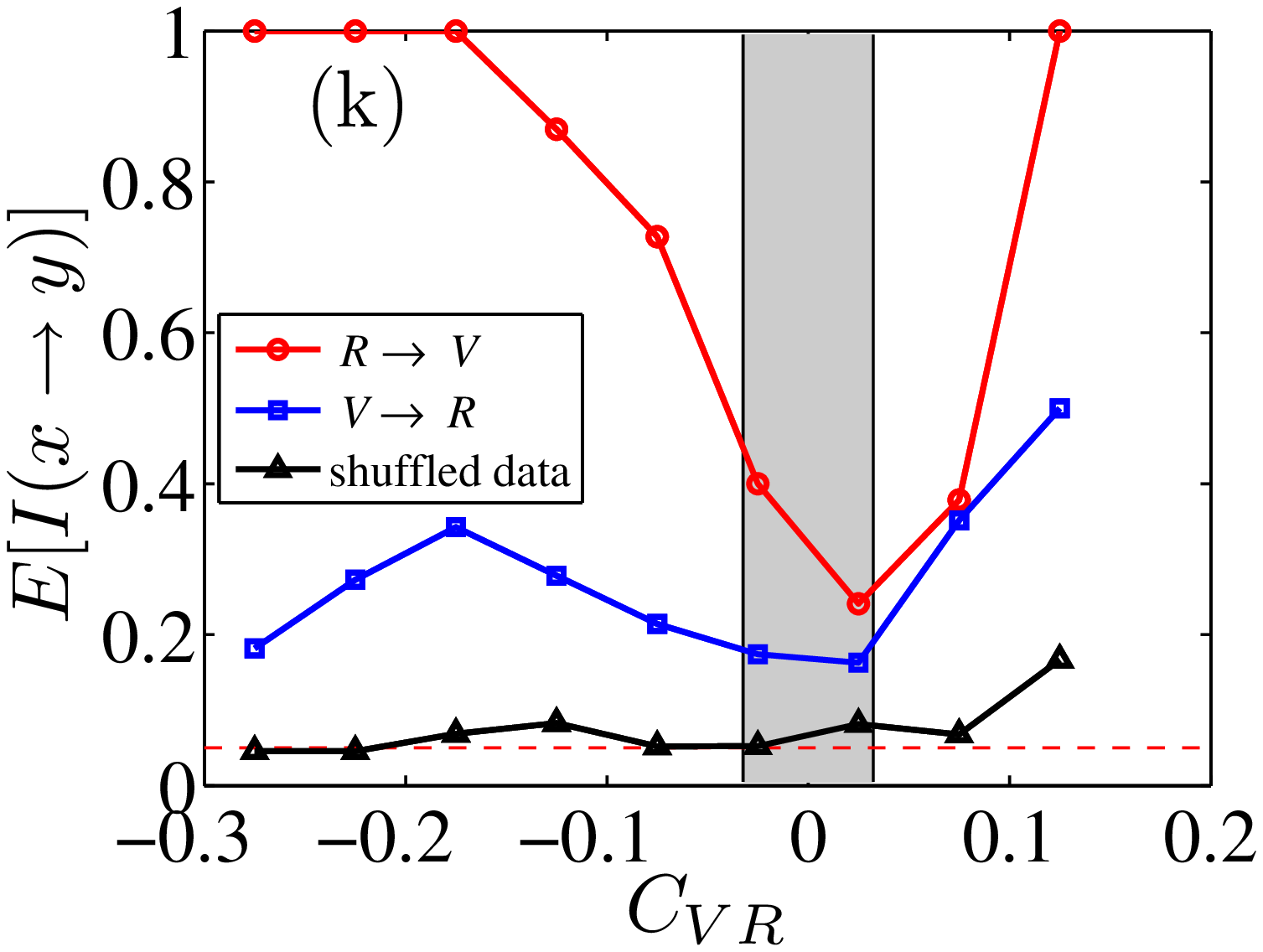}
\includegraphics[width=5.5cm]{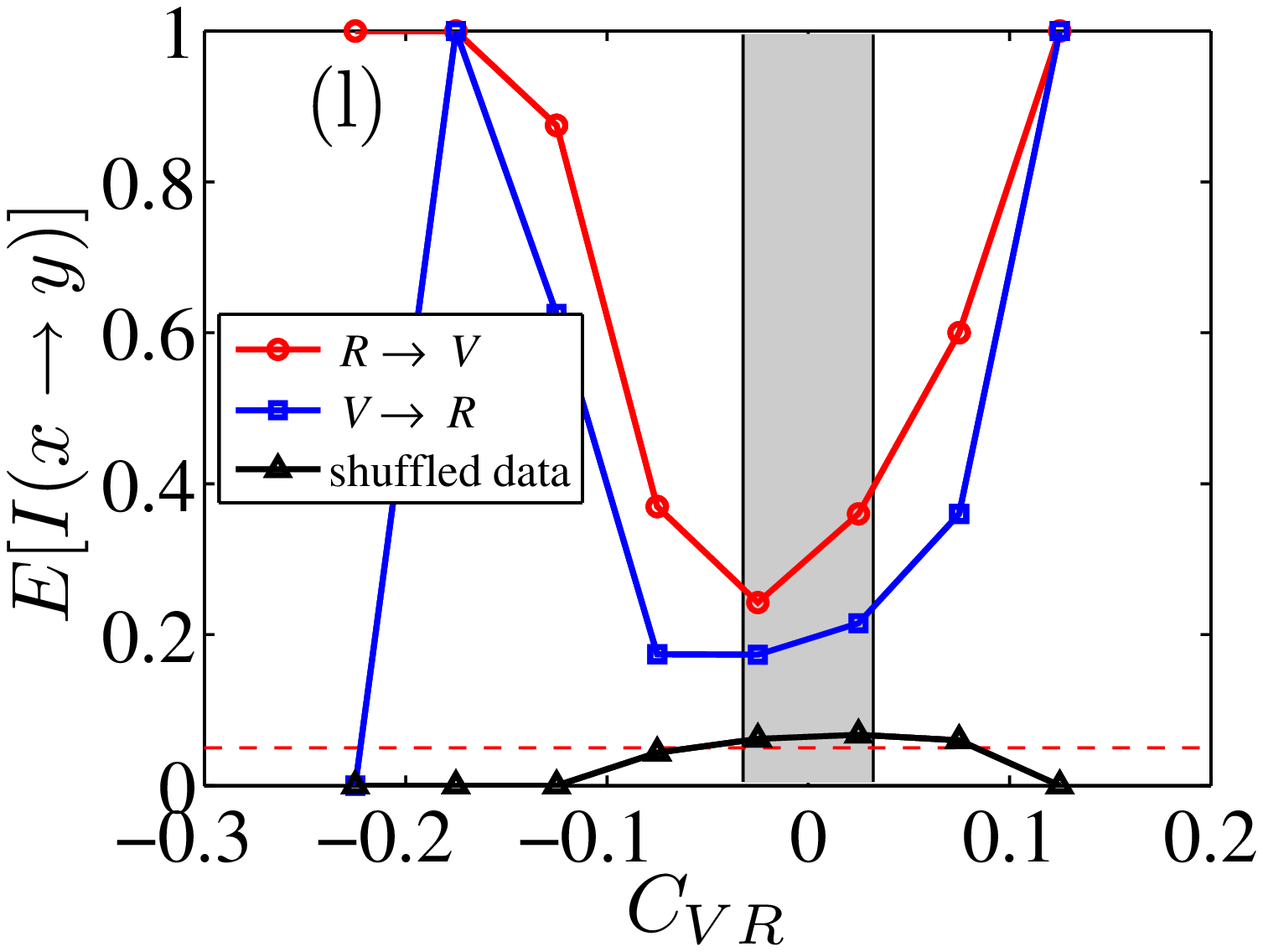}
 \caption{\label{Fig:Causality} (Color online) The first column (a,d,g,j) shows the results for all investors. The second column (b,e,h,k) shows the results for all individual investors. The third column (c,f,i,l) shows the results for all institutional investors. (a-c) Averaged autocorrelation functions $C_{V(t)V(t+\tau)}$ of the 15-min inventory variation $V$ for trending, reversing and uncategorized investors. The dashed lines give the 5\% significance level. (d-f) Averaged lagged cross-correlation functions $C_{V(t)R(t+\tau)}$ of the 15-min inventory variation for trending, reversing and uncategorized investors. The dashed lines bound the $\pm2\sigma$ significance interval. (g-i) Conditional expected value of the indicator $I(x \rightarrow y)$ of the rejection of the null hypothesis of non-Granger causality between $x$ and $y$ with $95\%$ confidence as a function of time horizons $\Delta T$. The dashed lines show the 5\% significance level. (j-l) Conditional expected value of the indicator $I(x \rightarrow y)$ as a function of the simultaneous cross-correlation $C[V_{i}(t),R(t)]$. The black symbols refer to the Granger test on shuffled data and the dashed lines bound $\pm2\sigma$ significance interval.}
\end{figure*}

We first investigate the autocorrelation function $C_{V(t)V(t+\tau)}$ of the inventory variation time series sampled in 15-min time intervals. Figure \ref{Fig:Causality}(a) shows the three autocorrelation functions for all the trending, reversing and uncategorized investors. Each autocorrelation function is obtained by averaging the autocorrelation functions of the investors in the same category to have better statistics. It is found that the inventory variation is long-term correlated and the correlation is significant over dozens of minutes, which can be partly explained by the order splitting behavior of large investors \cite{Vaglica-Lillo-Moro-Mantegna-2008-PRE,Moro-Vicente-Moyano-Gerig-Farmer-Vaglica-Lillo-Mantegna-2009-PRE,Vaglica-Lillo-Mantegna-2010-NJP}. We also find that the correlation is stronger among trending investors than reversing investors. Figure \ref{Fig:Causality}(b) and Fig.~\ref{Fig:Causality}(c) illustrate the results for individuals and institutions. We observe that institutions have stronger long memory than individuals. It implies that institutions are more specialized to their trading strategies than individuals \cite{Lillo-Moro-Vaglica-Mantegna-2008-NJP}.

Panels (d-f) of Fig.~\ref{Fig:Causality} illustrate the averaged lagged cross-correlation functions $C_{V(t)R(t+\tau)}$ between inventory variations and returns. The results in the three panels are qualitatively the same. For uncategorized investors, no significant cross-correlations are found between inventory variations and returns, which is trivial due to the ``definition'' of this category, as shown in Fig.~\ref{Fig:Categorization:test}(a-c). For trending and reversing investors, it is evident that the returns lead the inventory variations by dozens of minutes ($\tau<0$), where the cross-correlation $C_{V(t)R(t+\tau)}$ is significantly nonzero. When the price drops, trending investors will sell stock shares to reduce their inventory in a few minutes, while reversing investors will buy shares to increase their inventory. When the price rises, trending investors will buy shares to increase their inventory in a few minutes, while reversing investors will sell shares to reduce their inventory. In the meanwhile, we also observe nonzero cross-correlations for $\tau>0$ in shorter time periods, which means that the inventory variations lead returns.

To further explore the lead-lag structure between inventory variations and returns, we perform Granger causality analysis. We define an indicator $I(X\to Y)$, whose value is 1 if $X$ Granger causes $Y$ and 0 otherwise \cite{Lillo-Moro-Vaglica-Mantegna-2008-NJP}. In our analysis, the time resolution of the two time series is 15-min. The values of $I(V\to R)$ and  $I(R\to V)$ for all investors are determined at different time scales $\Delta{T}$. The average indicator values $E[I(V\to R)]$ and  $E[I(R\to V)]$ are plotted in Fig.~\ref{Fig:Causality}(g-i) with respect to $\Delta{T}$ for all investors, individual investors and institutional investors. Both $I(V\to R)$ and  $I(R\to V)$ are decreasing functions of $\Delta{T}$. We note that $E[I(X\to Y)]$ is the percentage of investors with $I(X\to Y)=1$. Figure \ref{Fig:Causality} shows that there are more investors with $I(R\to V)=1$ than investors with $I(V\to R)=1$. On average, bidirectional Granger causality is observed at the intraday time scales and the Granger causality disappears at the weekly level or longer. Moreover, individual investors are more probable to be influenced by the intraday price fluctuations than institutions, because the $I(R\to V)$ values of individuals are greater than those of institutions at the same time scale level.

We then investigate the impact of investor category on the causality indicator. The results for $\Delta{T}=4$ ({\it{i.e.}}, one hour) are depicted in Fig.~\ref{Fig:Causality}(j-l). The middle parts bounded by two vertical lines at $C_{VR}=\pm\sigma$ correspond to uncategorized investors. The left parts ($C_{VR}<-\sigma$) correspond to reversing investors and the right parts $C_{VR}>\pm\sigma$ correspond to trending investors. It is found that a investor adjusts his inventory following price fluctuations with very large probabilities when his $|C_{VR}|$ value is large. This conclusion holds for both individual and institutional investors. The strong Granger causality from inventory variations to returns and the weak but significant causality from returns to inventory variations cannot be attributed to the non-Gaussianity in the distributions of the variables, as verified by bootstrapping analysis. Qualitatively similar results are obtained for other $\Delta{T}$ values.

\subsection{Herding behavior}
\label{S2:herding}

Herding and positive feedbacks are essential for the boom of bubbles \cite{Sornette-2003,Sornette-2003-PR}. These topics have been studies extensively to understand the price formation process \cite{Lakonishok-Shleifer-Vishny-1992-JFE,Wermers-1999-JF,Nofsinger-Sias-1999-JF,Sias-2004-RFS}. Herding is a phenomena that a group of investors trading in the same direction over a period of time. Here, we try to investigate possible herding behaviors in different groups of investors.

\begin{table}[htb]
\centering
 \caption{\label{TB:Herding} Number of herding days for different groups of investors. The total number of trading days is 237. The superscripts ``$+$'' and ``$-$'' indicate buy herding and sell herding, respectively. The subscripts ``${\rm{d}}$'' and ``${\rm{s}}$'' indicate individuals and institutions, respectively. The time horizon is one day.}
\begin{tabular}{ccccccccccccccccccccccccccccccccccccccccccccccccccc}
  \hline\hline
       &&& \multicolumn{4}{c}{Reversing} &&& \multicolumn{4}{c}{Trending} &&& \multicolumn{4}{c}{Uncategorized} \\
       \cline{4-7}\cline{10-13}\cline{16-19}
            Code &&& $n_{\rm{d}}^{\rm{+}}$ & $n_{\rm{d}}^{\rm{-}}$ & $n_{\rm{s}}^{\rm{+}}$ & $n_{\rm{s}}^{\rm{-}}$
           &&& $n_{\rm{d}}^{\rm{+}}$ & $n_{\rm{d}}^{\rm{-}}$ & $n_{\rm{s}}^{\rm{+}}$ & $n_{\rm{s}}^{\rm{-}}$
           &&& $n_{\rm{d}}^{\rm{+}}$ & $n_{\rm{d}}^{\rm{-}}$ & $n_{\rm{s}}^{\rm{+}}$ & $n_{\rm{s}}^{\rm{-}}$
             \\\hline
     000001 &&& 60 & 63 & 0 & 0  &&& 0 & 0 & 0 & 0  &&& 6 & 8 & 0 & 0 \\
     000002 &&& 23 & 19 & 0 & 0  &&& 0 & 0 & 0 & 0  &&& 3 & 6 & 0 & 0 \\
     000012 &&& 26 & 37 & 0 & 0  &&& 0 & 0 & 0 & 0  &&& 2 & 1 & 0 & 0 \\
     000021 &&& 54 & 62 & 0 & 0  &&& 0 & 0 & 0 & 0  &&& 4 & 3 & 1 & 0 \\
     000063 &&& 34 & 34 & 2 & 1  &&& 0 & 0 & 0 & 0  &&& 0 & 0 & 1 & 1 \\
     000488 &&& 0 & 0 & 0 & 0  &&& 0 & 0 & 0 & 0  &&& 2 & 1 & 0 & 0 \\
     000550 &&& 17 & 34 & 0 & 0  &&& 0 & 0 & 0 & 0  &&& 0 & 0 & 2 & 1 \\
     000625 &&& 21 & 15 & 0 & 0  &&& 3 & 7 & 0 & 0  &&& 5 & 2 & 0 & 0 \\
     000800 &&& 29 & 25 & 0 & 0  &&& 0 & 0 & 0 & 0  &&& 7 & 7 & 0 & 0 \\
     000825 &&& 34 & 31 & 0 & 0  &&& 0 & 0 & 0 & 0  &&& 0 & 0 & 0 & 0 \\
     000839 &&& 61 & 58 & 0 & 0  &&& 0 & 0 & 0 & 0  &&& 7 & 7 & 0 & 0 \\
     000858 &&& 38 & 36 & 0 & 0  &&& 0 & 0 & 0 & 0  &&& 0 & 1 & 0 & 0 \\
     000898 &&& 55 & 40 & 0 & 0  &&& 0 & 0 & 0 & 0  &&& 3 & 4 & 0 & 0 \\
     200488 &&& 31 & 24 & 0 & 0  &&& 0 & 0 & 0 & 0  &&& 2 & 7 & 0 & 3 \\
     200625 &&& 6 & 6 & 0 & 0  &&& 0 & 0 & 0 & 0  &&& 3 & 6 & 0 & 3 \\
  \hline\hline
\end{tabular}
\end{table}

We study possible buy and sell herding behaviors among the same group of investors. Investors are classified into different groups based on their types (individual or institution) and their categories (reversing, trending or uncategorized). We define a herding index as follows \cite{Lillo-Moro-Vaglica-Mantegna-2008-NJP}:
\begin{equation}
 h = \frac{N^+}{N^++N^-},
 \label{Eq:HerdingIdx}
\end{equation}
where $N^+$ is the number of buying investors and $N^-$ is the number of selling investors in the same group over a given time horizon. When the herding index $h$ is smaller than 5\% under a binomial null hypothesis, we assess that herding is present. In our analysis, we fix the time horizon into one day and determine the number of days that herding was present for different groups of investors. The results are depicted in Table \ref{TB:Herding}.

According to Table \ref{TB:Herding}, there are no buying and selling herding days observed for trending institutions. For trending individuals and reversing institutions, herding is observed in only one stock on very few days. For categorized investors, we see slightly more herding days in a few stocks. For reversing investors, the number of herding days is greater than for other investors and we observe comparable buying and selling herding days. Our findings are consistent with those for the Spanish stock market, especially in the sense that reversing investors are more likely to herd \cite{Lillo-Moro-Vaglica-Mantegna-2008-NJP}. Our analysis also allows us to conclude that individuals are more likely to herd than institutions in 2003.

\section{Summary}
\label{S1:Summary}

In summary, we have studied the dynamics of investors' inventory variations. Our data set contains 15 stocks actively traded on the Shenzhen Stock Exchange in 2003 and the investors can be identified as either individuals or institutions.

We studied the cross-correlation matrix $C_{ij}$ of inventory variations of the most active individual and institutional investors. It is found that the distribution of cross-correlation coefficient $C_{ij}$ is asymmetric and has a power-law form in the bulk and exponential tails. The inventory variations exhibit stronger correlation when both investors are either individuals or institutions, which indicates that the trading behaviors are more similar within investors of the same type. The eigenvalue spectrum shows that the largest and the second largest eigenvalues of the correlation matrix cannot be explained by the random matrix theory and the components of the first eigenvector $u(\lambda_1)$ carry information about stock price fluctuation. In this respect, the behaviors differ for individual and institutional investors.

Based on the contemporaneous cross-correlation coefficients $C_{VR}$ between inventory variations and stock returns, we classified investors into three categories: trending investors who buy (sell) when stock price rises (falls), reversing investors who sell (buy) when stock price rises (falls), and uncategorized investors. We also observed that stock returns predict inventory variations. It is interesting to find that about 56\% individuals hold trending or reversing strategies and only 18\% institutions hold strategies. Moreover, there are far more reversing individuals (50\%) than trending individuals (6\%). In contrast, there are slightly more trending institutions (10\%) than reversing institutions (8\%). Hence, Chinese individual investors are prone to selling winning stocks and buying losing stocks, which provides supporting evidence that trading is hazardous to the wealth of individuals \cite{Odean-1998a-JF,Barber-Odean-2000-JF}.

\begin{acknowledgments}
 We are grateful to Fabrizio Lillo for fruitful discussion. This work was partially supported by the National Natural Science Foundation of China under grants 11075054 and 71131007, the Shanghai Rising Star Program under grant 11QH1400800, and the Fundamental Research Funds for the Central Universities.
\end{acknowledgments}

\bibliography{E:/Papers/Auxiliary/Bibliography}

\end{document}